\documentclass[aps,prb,twocolumn,floatfix,superscriptaddress]{revtex4-1}
\usepackage{comment}
\usepackage{caption}
\usepackage{subcaption}
\usepackage{graphicx}
\usepackage{amsmath}
\usepackage{amssymb}
\usepackage{slashed}
\usepackage{braket}
\usepackage{color}
\usepackage{multirow}
\usepackage[normalem]{ulem}
\usepackage[dvipsnames]{xcolor}

\newcommand{\be}{\begin{equation}}
\newcommand{\ee}{\end{equation}}

\begin{document}

\title{Theory of Competing Charge Density Wave, Kekule and Antiferromagnetic Fractional Quantum Hall States in graphene Aligned with boron nitride}
\author{Suraj S. Hegde}
\email{suraj.hegde@tu-dresden.de}
\affiliation{Institut für Theoretische physik, Technische Universität Dresden, 01069 Dresden, Germany}
\affiliation{Max-Planck institut für physik komplexer systeme, Nöthnitzer straße 38,01187 Dresden, Germany}

\author{Inti Sodemann Villadiego}
\email{sodemann@uni-leipzig.de}
\affiliation{Institut für Theoretische physik, Universität Leipzig, Brüderstraße 16, 04103, Leipzig, Germany}
\affiliation{Max-Planck institut für physik komplexer systeme, Nöthnitzer straße 38,01187 Dresden, Germany}

\begin{abstract}
    We investigate spin and valley symmetry-broken fractional quantum Hall phases within a formalism that naturally extends the paradigm of quantum Hall ferromagnetism from integer to fractional quantum Hall states, allowing us to construct detailed phase diagrams for a large class of multi-component states. Motivated by recent experiments on graphene aligned with a boron nitride substrate, we predict a sequence of transitions realized by increasing the magnetic field, starting from a sub-lattice polarized state to a valley coherent Kekule charge density wave state and further to an anti-ferromagnetic phase.  Moreover for filling fractions such as $\nu=\pm 1/3$, we predict that the system undergoes a transition at low fields, that not only differ by the spin-valley orientation of the fractionally filled flavors but also by their intrinsic fractional quantum Hall nature. This transition is from a Laughlin-like state to a two component Halperin-like state both with a charge density wave order. Moreover for $\nu=\pm 1/3,\pm 2/3$, we predict a ``canted Kekule density phase"(CaKD) where the spinors of integer and fractionally occupied components have different orientations in the valley Bloch sphere, in contrast to the Kekule state for the integer quantum Hall state at neutrality where both occupied components have the same orientation in the valley Bloch sphere.
\end{abstract}
\maketitle

\section{Introduction}

%{\bf reminder to cite: Si-Yu Li, Yu Zhang, Long-Jing Yin, and Lin He Phys. Rev. B 100, 085437 – Published 28 August 2019}

%{\bf reminder to cite: Veyrat, L. et al. Helical quantum Hall phase in graphene on SrTiO3. Science 367, 781–786 (2020).}

The quantum Hall regime in graphene offers a fantastic arena to investigate the interplay of broken symmetry, topology and fractionalization \cite{MacDonald90,Nomura06,Feldman09,Bolotin09,Goerbig11,Kharitonov12,Young_2012,Abanin13,young2014tunable,Sodemann14,Wu14,Wu15,Roy14,Peterson14,Amet15,Knothe15,Balram15,DaSilva16,Lukose16,Yan17,Polshyn18,Zibrov18,Stepanov_2018,ShaowenChen19,Maiti19,Schmitz20,Kim_2021,Atteia21,pierce2021,veyrat2020helical}, that continues to flourish thanks to a combination of high quality samples, enlarged degeneracy from its valley degree of freedom, and its exposed nature that allows new ways to measure and control the states of interest. For example, recent experiments have developed remarkable techniques to generate and detect long-range transmission of collective excitations in broken-symmetry integer and fractional quantum Hall states~\cite{Young_2012,Wei17,Wei18,Zibrov18,Stepanov_2018,aAssouline21,Fu_2021,pierce2021,Zhou2021}, and recent theoretical  and  scanning tunneling microscopy studies has provided detailed evidence of valence-bond quantum Hall Kekule-type states~\cite{liu2021visualizing,coissard2021,PhysRevB.100.085437,Das22}.

\begin{figure*}[]
\centering
\includegraphics[width=1.0\textwidth]{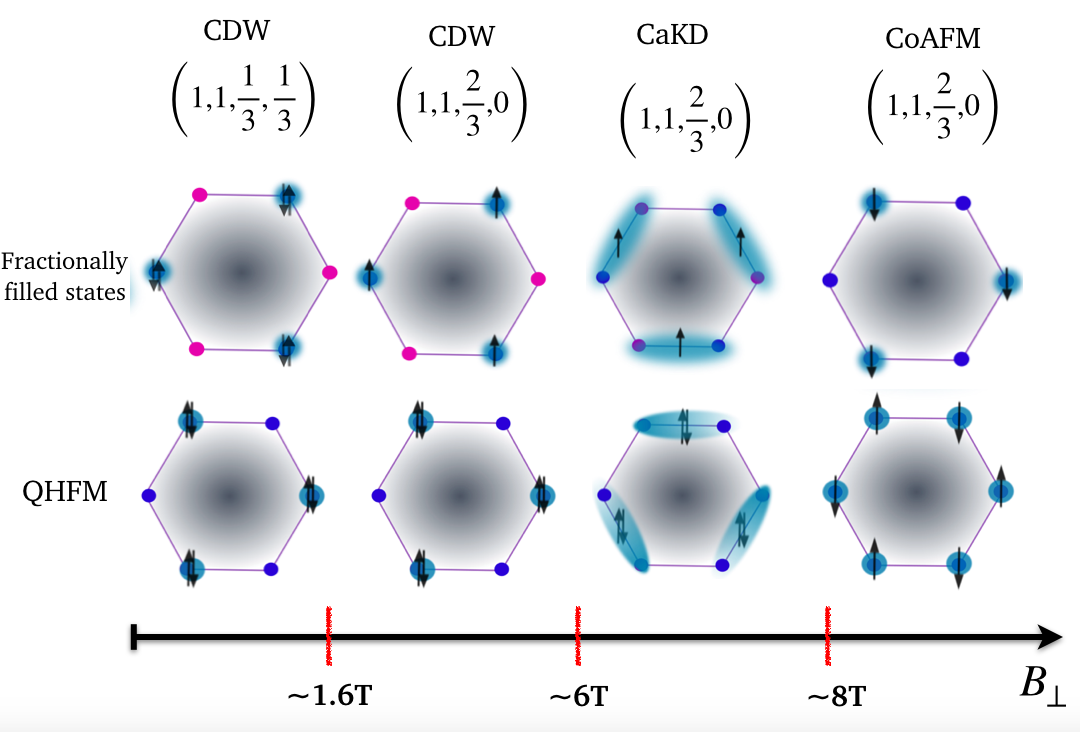} 
\caption{Sequence of transitions within  and between the valley-ordered phases and spin-ordered phases estimated using the variation theory for SU(4) symmetry breaking interaction in multi-component fractional quantum Hall states of graphene for a total filling $\nu=2/3$ measured relative to charge neutrality.}
\label{fig:AllTransitions}
\end{figure*}

Moreover, the possibility to align graphene with substrates, such as hexagonal-boron-nitride (hBN), can lead to a substantially enriched space of states. This is in fact one of the key purposes of this work, where we will demonstrate that a delicate competition of a variety of broken symmetry fractional quantum Hall states in the zeroth Landau level of graphene arise as a consequence of the interplay of the hBN-induced sub-lattice symmetry breaking and the tendency of the intrinsic interactions in graphene to stabilize anti-ferromagnetic states at high fields, as summarized in Fig.\ref{fig:AllTransitions}. 

Another key purpose of our work is to elaborate on a framework introduced by MacDonald and one of us in Ref.~\onlinecite{Sodemann14}, which generalizes the Hartree-Fock theory of integer quantum Hall ferromagnets to a vast class of experimentally relevant multi-component fractional quantum Hall states~\footnote{For a closely related discussion see also Ref.~\onlinecite{Abanin13}}. This framework allows to make detailed quantitative predictions of broken symmetry phase diagrams and quasi-particle excitation gaps, and, in particular, it can be applied to determine the competing phases arising in the model Hamiltonian introduced by Kharitonov~\cite{Kharitonov12}, which has proved to be a valuable tool for understanding the broken symmetry phases of the zeroth Landau level of graphene~\cite{Kharitonov12,Kharitonov12A,Sodemann14,Wu14,Knothe15,Murthy16,Tikhonov16,Lian16,Nova17,Lian17,Pientka17,Jolicoeur19}.  This model can be viewed as descending from the model introduced  by Aleiner, Kharzeev and Tsvelik~\cite{aleiner2007spontaneous} upon projection onto the zeroth Landau level of graphene~\cite{Kharitonov12}.

%See Fig.\ref{fig:AllTransitions} for a summary of the various broken valley- and spin- symmetry orders and the transitions between them.

While we will develop analytical formulae for a class of fractional quantum states with arbitrary fillings, we will focus specially on states realized at fillings $\nu=\pm 1/3,\pm 2/3$ which are often the most robust in experiments. At $\nu=\pm 1/3$ in particular we will describe a competition between a class of Laughlin-like states and singlet Halperin-like states that not only differ by the spin-valley orientation of the fractionally filled flavors but also by their intrinsic fractional quantum Hall nature, and are analogous to spin polarized and spin singlet states realized at total filling 2/3 in GaAs \cite{Eisenstein90,Lay97,Cho98,Nielmel00,Feldman09,Feldman12,Peterson15,YuheZhang16}.  This makes graphene an attractive platform to study these states, their edge modes and the interface between them, which due to its non-chiral nature, is a candidate to realize parafermion modes by coupling it to superconductors, which is an interesting platform for topological quantum computation\cite{Stern04,Bid10,Gross12,Gurman12,Mong14,Ronen18,Fabien19,Wang2021} .

Our paper is organized as follows: Sec.\ref{sec:Formalism} discusses the  framework that generalizes the theory of broken symmetry quantum Hall ferromagnetism to multi-component fractional quantum Hall states. The discussion here is kept general so as to apply to multi-component quantum Hall systems beyond graphene. Sec.\ref{sec:SU(4)Broken} then reviews how to apply this formalism to Kharitonov model of  symmetry breaking interactions in the $n=0$ Landau level of graphene, where $n=0$ corresponds to the orbital Landau level located at the Dirac point. Sec. \ref{sec:Neutrality}  then applies the formalism to compute the phase diagram of the broken symmetry states of the integer quantum Hall states at $\nu=0$ paying special attention to the modifications brought in by the hBN-induced sublattice potential. Sec. \ref{sec:fractional} generalizes these considerations to the phase diagrams of a large class of general fractional quantum Hall states.  Sec.\ref{sec:Transitions} presents our prediction of a series of transitions between valley- and spin-broken order in fractional quantum Hall states realised within the regime of experimental parameters. In Sec.\ref{sec:TopOrder}, we comment on the distinction in the universal topological order between the Laughlin-like and Halperin-like states. Finally, Sec.\ref{sec:Discussion} presents conclusions and discussions.

\section{Multi-component FQH states and variational theory of iso-spin symmetry-breaking short range interactions. }
\label{sec:Formalism}

\subsection{Constructing Correlated states on Quantum Hall Ferromagnet vacua}

We consider a multi-component quantum Hall system with $N$ internal flavors and imagine  that the Hamiltonian can be separated into a dominant part that is $SU(N)$ invariant plus smaller symmetry breaking perturbations. It is then natural to first understand many-body ground states of the $SU(N)$ invariant part, which appear in degenerate sets forming irreducible representations. Irreducible representations of $SU(N)$ arising from a system  $N_e$ electrons can be labeled by a set of $N$ integers satisfying:

\begin{equation}
n_1\geq n_2 ... \geq n_N \geq 0, \ \sum_i n_i=N_e.
\end{equation}
 
 These integers count the number of boxes in each row of a Young tableau depiction of the representation~\cite{pfeifer2003lie}, and, more intuitively, they simply count the invariant number of particles occupying each of the internal flavors~\footnote{For the familiar case of $SU(2)$ the $S^2$ quantum number is $S=(n_1-n_2)/2$.}. This leads to a generalization of the notion of filling factor into an $N$-component filling vector which specifies the $SU(N)$ invariant occupations of each of the flavors and is defined as:
 
\begin{equation}
(\nu_1, \nu_2, ... , \nu_N), \ \nu_i \equiv \frac{n_i}{N_\phi}, \ \sum_i \nu_i=\nu,
\end{equation}
where $N_{\phi}$ is the number of flux quanta.
When the fillings $\nu_i$ are integers the state corresponds to an integer quantum Hall ferromagnet (IQHFM) which are well-known to be exact ground states of a large class of repulsive $SU(N)$ invariant Hamiltonians projected into a Landau level. Moreover, IQHM states also serves as "perfect vacua on which one could construct correlated FQH states. Namely, if we have an eigenstate of the Hamiltonian with with $K \ (\leq N)$ fractionally filled components and $N-K$ empty components, labeled by $(\nu_1,..., \nu_K,0,....,0)$, then we can ``glue" this state with any IQHFM with $L \ (\leq N-K)$ integer filled components, to obtain a new eigenstate of the form $(1,...,1,\nu_1,..., \nu_K,0,...,0)$. The $SU(N)$ invariant interaction energy, denoted by $V$, of this new state can be obtained from that of the other states, and is given by \cite{IntiThesis} :

\begin{equation}\label{Epsi1}
V[\Psi_{L+K}]=V[\Psi_{K}]+V[\Psi_{L}]+\nu_K  L N_\phi V_H,
\end{equation}

%\be\label{Epsi1}
%V[\Psi_{L+K}]=V[\Psi_{K}]+V[\Psi_{L}]+\nu_K  L \frac{N_\phi}{2\pi l^2} \int d^2r \ v(r),
%\ee

\begin{equation}
V_H = \frac{1}{2\pi l^2} \int d^2r \ v(r),
\end{equation}

\noindent where $\Psi_{K}$ denotes the $K$-component FQH state with filling $\nu_K=\sum_{i=1}^{K}\nu_i$, $\Psi_{L}$ denotes the $L$-component IQHFM state, and $\Psi_{K+L}$ denotes the glued state of these two. $v(r)$ is simply the Hartree potential between the electrons in the fractionally and integer occupied components, and would be absent if a neutralizing background was present, as it is the case with Coulomb interactions. $l$ is the magnetic length.  

The construction described above is based on adding correlated {\it particles} to the IQHFM. It is also possible to construct a different set of states by adding correlated {\it holes}. Namely, for any eigenstate with fillings $(\nu_1,..., \nu_K,0,....,0)$ one can obtain other correlated states in which these particles are removed from an IQHFM with $L \ ( \geq K)$ integer filled components, leading to a new state with $L-K$ integer filled components and the remaining fractionally filled as follows $(1,...1,1-\nu_K,\cdots,1-\nu_1,0,\cdots,0)$. The energy of these states can be shown to be related as follows\cite{IntiThesis} :

%\begin{equation}\label{Epsi2}
%V[\Psi_{L-K}]=V[\Psi_K]+N_\phi (L-\nu)(L V_H-\int \frac{d^2q}{(2\pi)^2} v(q) |F(q)|^2),
%\end{equation}

\begin{align}
    V[\Psi_{L-K}]=&V[\Psi_K] \nonumber \\ &+N_\phi (L-\nu)\bigg(L V_H-\int \frac{d^2q}{(2\pi)^2} v(q) |F(q)|^2 \bigg),
\end{align}
 where $|F(q)|^2=\exp(-q^2 l^2/2)$ is the squared density form factor of the lowest Landau level and $v(q)$ is the Fourier transform of the Hartree potential $v(r)$.

\subsection{Hard-core states and variational theory of short range SU(N) breaking interactions}\label{Hardcore}

The above two rules allow to construct a large set of new FQHE states and determine their energy from knowledge of FQHE states constructed on vacuum. An important set of FQHE states in vacuum are those we will refer to as ``hard-core" states. A hard-core state is one in which the wave-function vanishes when any two-particles approach each other regardless of whether they have the same $SU(N)$ pseudo-spin flavor. Clearly this property endows these hard-core states with energetically desirable correlations for repulsive interactions, and this set includes many celebrated wave-functions like the Laughlin states, many of the Halperin multicomponent states, the Jain states, the Moore-Read states and others.

%\begin{equation}\label{}
%V[\Psi_{L-K}]=V[\Psi_K]+N_\phi (L/2-\nu)(L V_H-V_X),
%\end{equation}

An important property of the set of any hard-core states is that they necessarily have total filling factors restricted to $\nu \leq 1$~\cite{Sodemann14,IntiThesis}. This can be seen easily by noting that any wave-function which vanishes whenever two particles approach each other must necessarily be proportional to the Laughlin-Jastrow-like function $\prod_{i<j} (z_i-z_j) $ with the indices $i,j$ running over all the particles times another analytic polynomial of the coordinates. But the wavefunction associated with the Laughlin-Jastrow-like factor has filling factor 1, and since multiplication by an analytic polynomial always reduces the density of the wave-function, it follows that their fillings are restricted to $\nu \leq 1$.

Another immediate property of hard-core states is that they are zero eigenstates of any generic delta function interaction even if it breaks the $SU(N)$ symmetry. To illustrate this, let us consider the following $SU(N)$ symmetry breaking Hamiltonian:

\begin{align}\label{Eq:Ha}
    H_a = \sum_{i<j, \alpha\beta} g_{\alpha\beta} \mathcal{T}_{\alpha}^i  \mathcal{T}_{\beta}^j \delta^{(2)}({\bf r}_i-{\bf r}_j) - \sum_{i\alpha} h_{\alpha} \mathcal{T}_{\alpha}^i,
 \end{align}

\noindent here $\mathcal{T}_{\alpha}$ are generators of $SU(N)$ and $g_{\alpha\beta}$ is a symmetric matrix, and $h_\alpha$ denote the strength of uniform single-particle symmetry-breaking terms. It is easy to show that the hard-core states are exact ground states of the above Hamiltonian, simply because they have zero probability for any two particles to coincide at the same point in space regardless of their $SU(N)$ flavor, and therefore they are annihilated by the delta function in Eq.\ref{Eq:Ha}. Moreover the $SU(N)$ flavors of the occupied states can be chosen to diagonalize the single-particle matrix $h_\alpha \mathcal{T}_{\alpha}$, so that they are eigenstates of the full Hamiltonian in Eq.\ref{Eq:Ha}. 

Now the states with $\nu>1$ that follow from the two previously discussed rules are not necessarily exact eigenstates of this Hamiltonian. Let us imagine, however, that the $SU(N)$ invariant part of the Hamiltonian is much stronger than the symmetry breaking terms in Eq.\ref{Eq:Ha}, and therefore that the orbital two-particle correlation functions of these states are sufficiently rigid so as to remain well approximated by the ideal $SU(N)$ invariant states, and that the role of the terms in Eq.\ref{Eq:Ha} is to select the ground state within the $SU(N)$ degenerate manifold. This assumption allows to construct a variational mean field theory of these states somewhat analogous to Hartree-Fock theory, albeit extended to states that are far from being approximated as Slater determinants, namely, the FQH states generated by the two previously discussed rules. Specifically, consider an IQHFM vacuum occupying $L $ flavors, which can be parametrized by a projector of the form:

\be\label{Pi}
P_i= \sum_{k=1}^L \ket{\chi_k} \bra{\chi_k},
\ee

and consider a hard-core FQH state of $K$ flavors, and to which we associate a density matrix of fractional fillings associated as follows:

\be\label{Pf}
P_f= \sum_{k=1}^K \nu_{k}\ket{\chi_{L+k}} \bra{\chi_{L+k}}.
\ee

In writing $P_i$ and $P_f$ we are assuming that all the spinors $\ket{\chi_k},\ket{\chi_{L+k}}$ are orthogonal and normalized. Now, it can be shown \cite{Sodemann14,IntiThesis} that the expectation value of the symmetry breaking Hamiltonian in Eq.\ref{Eq:Ha} for the ``glued" state of the above IQHFM and FQH states can be written as:

\begin{align}\label{Haexp}
    \frac{\langle H_a \rangle}{N_\phi} = \frac{1}{2} \text{Tr}(P_i H_i^{HF})+\text{Tr} (P_f H_i^{HF})  \nonumber\\
    -\sum_{\alpha} h_{\alpha} \text{Tr}((P_i+P_f)\mathcal{T}_{\alpha}). 
\end{align}

\noindent where $H_i^{HF}$ is the mean-field Hartree-Fock Hamiltonian arising from interactions with the particles occupying integer filled components of the IQHFM, and it is given by:   

\begin{align}
    H_i^{HF}= \sum_{\alpha,\beta}u_{\alpha\beta}[\text{Tr}(P_i \mathcal{T}_{\alpha}) \mathcal{T}_{\beta} -\mathcal{T}_{\alpha} P_i \mathcal{T}_{\beta}]
\end{align}
where $u_{\alpha,\beta} = \frac{g_{\alpha,\beta}}{2 \pi l^2}$.  Notice that when the fractionally filled components are empty (i.e. $P_f=0$), the energy in Eq.\eqref{Haexp} reduces to the Hartree-Fock energy of the integer quantum Hall ferromagnetic states (compare e.g. with Eqs.(44)-(48) of Ref.~\onlinecite{Kharitonov12}). Thus we see that our description is a natural extension of the theory of integer quantum Hall ferromagnets to the case of multicomponent fractional quantum Hall states.

Therefore the task is reduced to minimizing the energy from Eq.~\eqref{Haexp} as a function of the spinors $\ket{\chi_k}$ parametrizing the IQHFM from Eqs.\eqref{Pi} and \eqref{Pf}. The above equations are obtained from the assumption of hard-core repulsion and the following property of completely filled spinors 
\begin{equation}
    \hat{\rho}_m(r) \ket{\Psi}=\frac{1}{2\pi \ell^2} \ket{\Psi},
\end{equation}
where $\hat{\rho}_m(r)=P_{LLL}\sum_i (\delta(r-r_i) \ket{\chi_m}_{ii}\bra{\chi_m})P_{LLL}$ is the density projected to the mth completely filled spinor. We would like to also mention in passing, that while the above framework captures a large variety of multi-component states, other multi-component states that deviate from this picture, to this date have not been understood theoretically, have been found in numerical studies~\cite{Wu15,Le2021}.

%The completely filled spinor acts as the variational parameter in the Hartree-Fock theory.

\section{Weakly broken SU(4) FQH states in graphene on h-BN}
\label{sec:SU(4)Broken}
In this section we apply the formalism of Sec.\ref{sec:Formalism} to the specific case of SU(4) symmetry breaking interactions known to occur in FQHEs of graphene. 
The integer and fractional quantum Hall states of graphene under high magnetic fields have been considered in detail in various works \cite{Kharitonov12,Abanin13,Sodemann14} and in particular recent experiments have developed ingenious ways to probe the pseudo-spin order in these systems by exploiting the coupling of their order parameters to non-local transport phenomena\cite{Stepanov_2018,Zibrov18,Wei18,Polshyn18,Zhou2021}.

We focus on the zero Landau level of graphene, which has N=4 flavors arising from two valleys and two spins. The long-range part of the Coulomb interactions projected into these Landau levels are viewed as the dominant SU(4) invariant part of the Hamiltonian.  In addition there are short-distance lattice scale corrections to the Coulomb interactions~\cite{Kharitonov12,aleiner2007spontaneous} that can be modeled as sub-lattice and valley dependence contact interactions~\footnote{For counterparts of these interactions in bilayer graphene see e.g. Refs.~\cite{lemonik2010spontaneous,lemonik2012competing}. Including these terms is crucial because they play the role of the leading symmetry breaking terms in the Landau level, and therefore determine the specific broken symmetry ground states.} We also imagine that the alignment of graphene with an hBN substrate gives rise to a single-particle valley splitting with strength $\Delta$, as relevant for experiments in Ref.\onlinecite{Zibrov18}. Together with the ordinary Zeeman term and short distance interactions considered in previous studies \cite{Kharitonov12,Sodemann14,IntiThesis} the Hamiltonian in Eq. \ref{Eq:Ha} can be written as: 

%\begin{align}
    %H_a = \sum_{i<j, \alpha} u_{\alpha} \mathcal{T}_{\alpha}^i \delta^{(2)}({\bf r}_i-{\bf r}_j) \mathcal{T}_{\alpha}^j - \sum_i h_{\alpha} \mathcal{T}_{\alpha}^i,
    %\end{align}

\begin{align}
    H_a = \sum_{i<j, \alpha}( g_{\perp}( \tau_{x}^i  \tau_{x}^j + \tau_{y}^i \tau_{y}^j)+g_z\tau_{z}^i \tau_{z}^j )\delta^{(2)}({\bf r}_i-{\bf r}_j) \nonumber \\ - \sum_i( h \sigma_z^i + \Delta \tau_z^i)
\end{align}

\noindent where $\tau_i$ and $\sigma_i$ are Pauli matrices for the valley iso-spin and intrinsic spin respectively, $g_{\perp,z}$ are the strengths of the delta-function interactions,  $h$ is the spin-Zeeman term and $\Delta$ is the valley-Zeeman term. The valley-Zeeman term comes from sublattices A and B having different energies, originating from the alignment of graphene with an hBN substrate, as seen in penetration capacitance measurements~\cite{Zibrov18}. These terms break the SU(4) weakly in the sense that $g_{\perp},g_z< e^2/\epsilon \ell$ and $h,\Delta < e^2/\epsilon \ell$. The values of $g_\perp,g_z$ are dependant on the magnetic field component perpendicular to the graphene plane $B_{\perp}$ ($g_{\perp,z}\propto B_{\perp}$), whereas the Zeeman term $h$ is determined by the total magnetic field. From the above short-range interaction terms one can define two energy scales:

\begin{equation}
u_{z,\perp} = \frac{g_{z,\perp}}{2 \pi l^2}
\end{equation}

%This term plays a crucial role in our analysis and in relating certain the results to the experiments. 

%Thus, they determine the valley or spin order realised on these fractional quantum Hall states.

%Their competition can be tuned by tilting the magnetic field with respect the graphene plane. 

%The term parametrised by $g_z$ preserves the valley of the scattering electrons and has opposite signs for intra- and inter-valley interations. The $g_{\perp}$ is an inter-valley scattering term and preserves the number of electrons in each valley. 

Now, let us obtain the contribution of these anisotropy energies in order to determine the different phases realised in the parameter space, following the formalism described in Sec.~\ref{Hardcore}. The many-body states we consider are the ones which are constructed from the hardcore states which have a filling of the form: $(1,1,\nu_3,\nu_4)$~\footnote{There are more possibilities of the form $(1,\nu_2,\nu_3,0)$, but as discussed in Ref.\onlinecite{Sodemann14}, these states are disfavored much more strongly by the short range interactions and tend to be stabilized at very small magnetic fields, and therefore we will not consider them here. This is also supported by recent detailed numerical studies~\cite{Le2021}}. The variational coherent states associated with the completely filled flavors are denoted by $\{ \ket{\chi_1},\ket{\chi_2} \}$. The ones of the partially filled flavors will be denoted by $\{ \ket{\chi_{3}},\ket{\chi_{4}} \}$. The weighed projectors onto these states are denoted by $P_i=\ket{\chi_1}\bra{\chi_1}+ \ket{\chi_2}\bra{\chi_2}$ and $P_f= \nu_3 \ket{\chi_{3}}\bra{\chi_{3}}+ \nu_4\ket{\chi_{4}}\bra{\chi_{4}} $. The expectation value of the anisotropy energy per flux quantum is given by\cite{Sodemann14,IntiThesis}
\begin{align}
\label{Eq:Anisotropy}
    \frac{\langle H_a \rangle}{N_\phi} = \frac{1}{2} \text{Tr}(P_i H_i^{HF})+\text{Tr} (P_f H_i^{HF}) \nonumber \\-\frac{h}{2}\text{Tr} (P_i \sigma_z)-\frac{\Delta}{2}\text{Tr} (P_i \tau_z) \nonumber \\ 
    H_i^{HF}= \sum_{\alpha}u_{\alpha}[\text{Tr}(P_i \tau_\alpha) \tau_\alpha -\tau_\alpha P_i \tau_\alpha] -h \sigma_z-\Delta \tau_z \end{align}

In the subsequent sections we will minimize the above variational energy with respect to the SU(4) orientation of the spinors $\{ \ket{\chi_1},\ket{\chi_2},\ket{\chi_3},\ket{\chi_4} \}$, for various cases.

 %This formalism has been used extensively to predict various fractional Hall states realised in graphene and to obtain the detailed phase diagram in these states\cite{Sodemann14,Kharitonov12}. 

\section{Phase diagram for IQH states at neutrality with valley-Zeeman term }
\label{sec:Neutrality}
 
  \begin{figure}[]
\centering
\includegraphics[width=.4\textwidth]{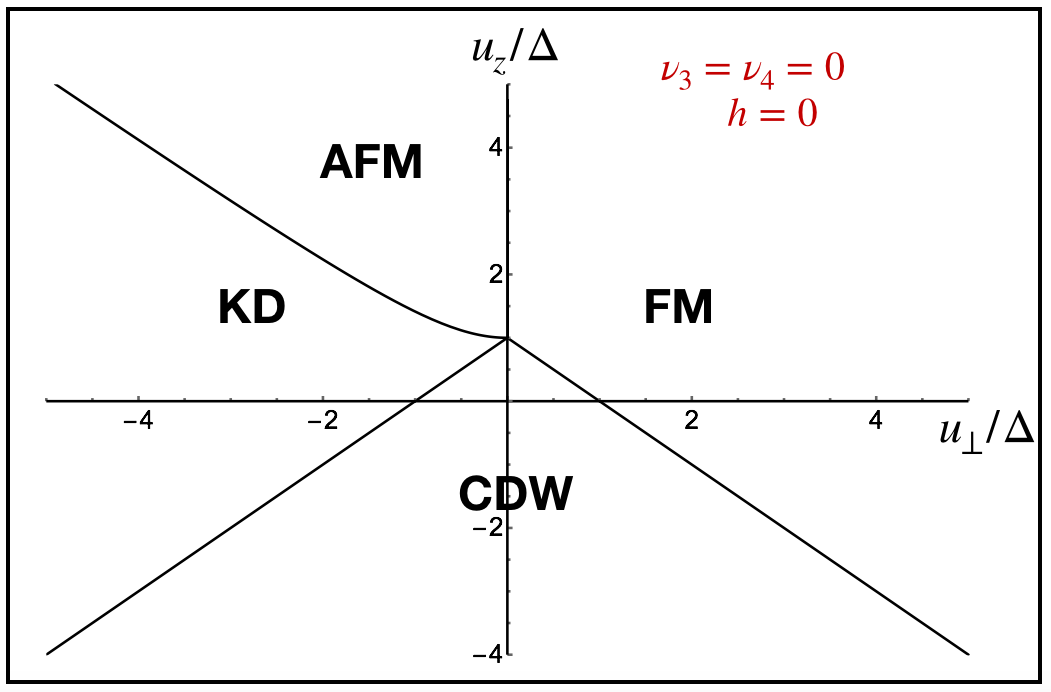}
\caption{Phase diagram of $(1,1,0,0)$ integer Quantum Hall ferromagnetic state in graphene in the presence of valley-Zeeman term $\Delta$ (sub-lattice staggered potential).}
 \label{fig:PhaseDiagram(1,1)}
\end{figure}

In this section we will consider the symmetry broken phases and phase diagram of the integer quantum Hall effect of graphene at neutrality.  We sometimes will denote the total filling by $\tilde{\nu}$, such that $\tilde{\nu} \in [0,4]$.  Neutrality corresponds to $\tilde{\nu}=2$, and this relates to the other frequently used notation of filling $\nu \in (-2,2)$ as follows: $\tilde{\nu}=2+\nu$. 

%Notice that this way of labeling filling fractions differs from the one 

The phase diagram of $\tilde{\nu}=2$ ($\nu=0$) quantum Hall state in graphene at neutrality has been studied extensively\cite{Kharitonov12,Abanin13,Wu14,Zibrov18}. Here we will review relevant aspects of it in order  set-up the notation and to include the valley-Zeeman term induced by the hBN substrate in the same manner as in Ref. \onlinecite{Zibrov18}. This state also serves as the vacuum for creating fractionally filled phases that we discuss in the next section. Charge neutrality corresponds to occupation of two components $(\nu_1,\nu_2,\nu_3,\nu_4)=(1,1,0,0)$. 

Within the theory described in Sec.\ref{sec:SU(4)Broken}, the possible ground states are parametrized by the projector: $P_i= \ket{\chi_1}\bra{\chi_1}+ \ket{\chi_2}\bra{\chi_2}$. We will also further restrict to states in which the occupied spinors do not have valley-spin entanglement\cite{Kharitonov12,Sodemann14}, namely states for which the SU(4) spinors can be taken as a tensor product of the state in the valley space times one in the spin space~\footnote{An interesting study found spin-valley entanglement near the egde of the integer quantum Hall ferromagnets~\cite{Knothe15}, but here we are focusing on bulk ground states.}. Restriction to this sub-set is parametrised by the projector $P_i$:
\begin{equation}
\label{Eq:Projector}
    P_i=P_{n_1} \otimes P_{s_1} + P_{n_2} \otimes P_{s_2},
\end{equation}
where $\vec{n}_1,\vec{n}_2$ are the unit vectors in the Bloch sphere of the valley iso-spin and $\vec{s}_1,\vec{s}_2$ are the unit vectors in the spin space. $P_{n_{j}}$ is the valley-spin density matrix given by
 \begin{equation}
     P_{nj}= \ket{\vec{n}_{j}}\bra{\vec{n}_{j}}= \frac{1}{2}(1+\vec{n}_{j} \cdot \vec{\tau})
 \end{equation}
 and of similar structure for the spins $\vec{s}_1,\vec{s}_2$. In addition one needs to enforce that the two projectors in the right hand side of Eq.\eqref{Eq:Projector} are orthogonal. This is achieved by either making the spin vectors anti-parallel while leaving their valley orientations arbitrary, and we refer to this states ``valley ordered", or by the converse choice in which the valley vectors are anti-aligned and the spin vectors are arbitrary, and we refer to the latter as ``spin ordered" states. The valley ordered states can be written as:
 \begin{eqnarray}
     \chi_1 =\ket{\vec{n}_1 }\otimes \ket{\vec{s}},\quad \chi_2 =\ket{\vec{n}_2}\otimes \ket{-\vec{s}}, \nonumber \\
     P_n =P_{n_1} \otimes P_s + P_{n_2} \otimes P_{-s}
     \label{Eq:Pn}
 \end{eqnarray}
 These contain states that correspond to charge density wave (CDW) and Kekule-distortion (KD) phases. The spin-ordered states are: 
  \begin{eqnarray}
     \chi_a =\ket{\vec{n}} \otimes \ket{\vec{s}_1},\quad \chi_b =\ket{-\vec{n}}\otimes \ket{\vec{s}_2},  \nonumber \\
     P_s =P_{n} \otimes P_{s_1} + P_{-n} \otimes P_{s_2}
     \label{Eq:Ps}
 \end{eqnarray}
 These contain states such as ferromagnetic (FM) phase and various anti-ferromagnetic (AFM) phases.
%   \begin{eqnarray}
%      \chi_a =\ket{\vec{n}} \otimes \ket{\vec{s}},\quad \chi_b =\ket{-\vec{n}}\otimes \ket{-\vec{s}}, \\
%      P_{ns} =P_{n} \otimes P_{s} + P_{-n} \otimes P_{-s}
%  \end{eqnarray}
From Eq.\eqref{Eq:Anisotropy} the anisotropy energy of the valley-ordered states parametrized in Eq.\eqref{Eq:Pn} can be found to be:
 \begin{equation}
     E_{va}= u_{\perp}(n_{1x}n_{2x}+n_{1y}n_{2y})+u_z n_{1z}n_{2z}-\Delta (n_{1z}+n_{2z})
     \label{Eq:EvaNeutral}
 \end{equation}
 
 Similarly, the anisotropy energy of the spin-ordered states parametrised in Eq. \eqref{Eq:Ps} is found to be:
 \begin{equation}
     E_{sa} = -u_{\perp}(1+ \vec{s}_1. \vec{s}_2)-u_z -h(s_{1z}+s_{2z})
     \label{Eq:EsaNeutral}
 \end{equation}
 
\begin{table}
\centering
\renewcommand{\arraystretch}{1.3}
\begin{tabular}{|c| c| c| c|}
\hline %inserts horizontal line
State & Valley & Spin & Energy  \\[0.5ex] 
\hline 
CDW & $n_{1z}=n_{2z}=1$ & $\vec{s}_1=-\vec{s}_2$ & $u_z-2\Delta$ \\    
\hline 
Kekule & $n_{1z}=n_{2z}=\frac{\Delta}{u_{\perp}-u_z}$ & $\vec{s}_1=-\vec{s}_2$ & $u_{\perp}-\frac{\Delta^2}{u_z-u_{\perp}}$  \\ 
\hline 
CaAFM & $\vec{n}_1=-\vec{n}_2$ & $s_{1z}=s_{2z}=\frac{h}{2u_{\perp}}$  & $-u_z-\frac{h^2}{2|u_{\perp}|}$  \\ 
\hline
FM & $\vec{n}_1=-\vec{n}_2$ & $ \vec{s}_1=\vec{s}_2$ & $-2u_{\perp}-u_z-2h$  \\ 
\hline 
 
\end{tabular}
\caption{Spin-valley orientations of integer quantum Hall ferromagnets at neutrality and energies with Zeeman and valley-Zeeman coupling. }
\label{tsymm}
%\label{tsymm}  
\end{table} 
 
 %At neutrality $\vec{n}_3=\vec{n}_4=0, \vec{s}_3=\vec{s}_4=0$
 
 Minimization of these energy functionals leads to four phases listed in Table \ref{tsymm} and to the corresponding phase diagram shown in Fig.\ref{fig:PhaseDiagram(1,1)}.

\section{Phase diagram of FQHE states in graphene with valley-Zeeman term}
\label{sec:fractional}

\begin{table*}
\renewcommand{\arraystretch}{1.3}
\begin{tabular}{|c|c|c|l|}
\hline %inserts horizontal line
State & Valley iso-spin & Spin & Energy  \\[0.5ex] 
\hline 
CDW & $n_{1z}=-n_{3z}=n_{2z}=-n_{4z}=1$ & $\vec{s}_1=-\vec{s}_2=\vec{s}_3=-\vec{s}_4$ & $u_z(1-2\nu) -2\nu u_{\perp}-\Delta(2-\nu)-h(\nu_3-\nu_4)$  \\ 
\hline 
CantedKD & $\vec{n}_{1}=-\vec{n}_{3}$, $\vec{n}_{2}=-\vec{n}_{4}$,  & $\vec{s}_1=-\vec{s}_2=\vec{s}_3=-\vec{s}_4$ & $(1-\nu) u_{\perp}\sqrt{\frac{(1-\nu)^2(u_z^2-u_{\perp}^2)-(1-\nu_4)^2\Delta^2}{(1-\nu)^2(u_z^2-u_{\perp}^2)-(1-\nu_3)^2\Delta^2}} \bigg( 1-\frac{\Delta^2 (1-\nu_3)^2}{(1-\nu)^2(u_z^2-u_{\perp}^2)} \bigg)$ \\
     & $n_{3z}\neq n_{4z}$ & & $- \frac{(1-\nu_3)(1-\nu_4)u_z \Delta^2}{(1-\nu)(u_z^2-u_{\perp}^2)}-2\nu  u_{\perp}-\nu u_z$  \\ 
   \hline 
CoAFM & $\vec{n}_1=-\vec{n}_2=\vec{n}_3=-\vec{n}_4$  & $s_{1z}=-s_{2z}=-s_{3z}=s_{4z}$ & $-u_z-2\nu u_{\perp}-h(\nu_3-\nu_4)-\Delta(\nu_3-\nu_4)$  \\ 
\hline
CaAFM & $\vec{n}_1=-\vec{n}_2=\vec{n}_3=-\vec{n}_4$ &$\vec{s}_1=-\vec{s}_3$,$\vec{s}_2=-\vec{s}_4=1$, & $|u_{\perp}|(1+\nu)-\frac{|u_{\perp}|(1-\nu)}{2}\bigg( \frac{1-\nu_4}{1-\nu_3} +\frac{1-\nu_3}{1-\nu_4} \bigg) -\frac{h^2}{2|u_{\perp}|} \frac{(1-\nu_3)(1-\nu_4)}{(1-\nu)}$   \\
 & &$s_{3z}\neq s_{4z}$ & $-u_z -\Delta(\nu_3-\nu_4)$ \\
\hline
FM & $\vec{n}_1=-\vec{n}_2=\vec{n}_3=-\vec{n}_4$ & $s_{1z}=s_{2z}=s_{3z}=s_{4z}=1$ & $-u_z -2u_{\perp}-h(2-\nu)-\Delta(\nu_3-\nu_4)$  \\ 
\hline 
\end{tabular}
\caption{ Orientations of valley iso-spin and spin for different phases at fractional filling $(1,1,\nu_3,\nu_4)$ and the corresponding energies. For the expressions of $n_{3z},n_{4z}$ see \ref{Eq:OptimalValeySpin}. The expression for $s_{3z},s_{4z}$ are given in Ref.\onlinecite{Sodemann14}.
\label{frac-tsymm}}
\end{table*}

\begin{figure}[]
\centering
\includegraphics[width=.4\textwidth]{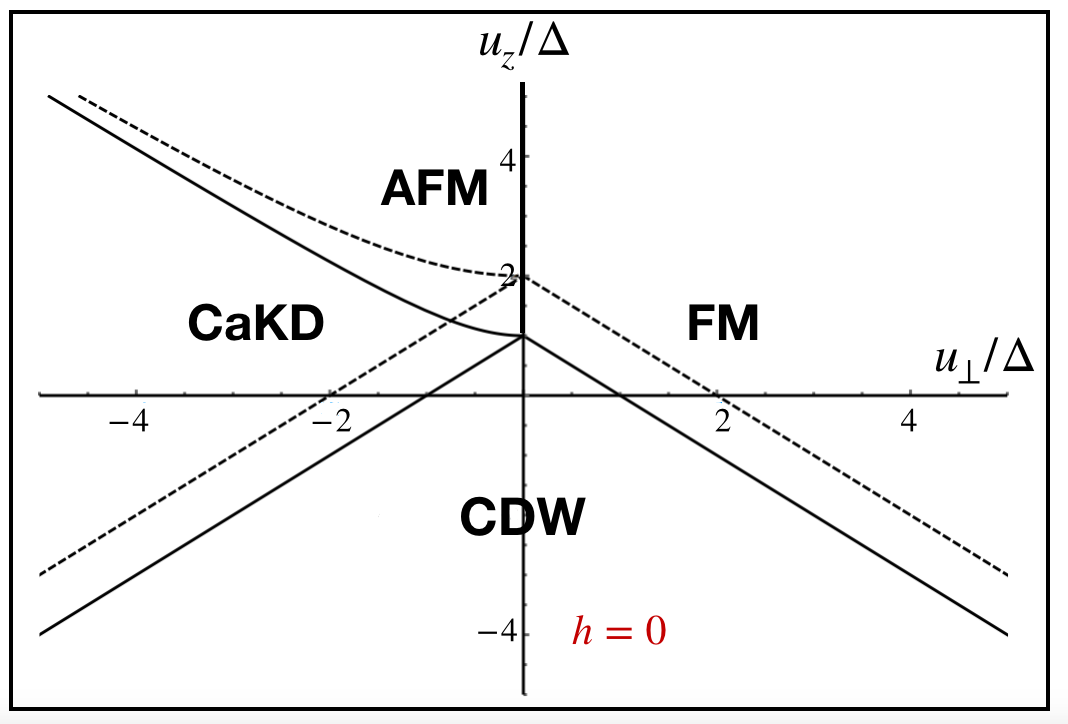}
\caption{Phase diagrams in the presence of valley-Zeeman term $\Delta$ (sub-lattice staggered potential). The phase boundaries of $(1,1,2/3,0)$ state are indicated by dark lines and phase boundaries of $(1,1,1/3,1/3)$ indicated by dashed lines. }
 \label{fig:PhaseDiagram}
\end{figure}

In this section, we will obtain the phase diagram for a class of fractional quantum Hall phases in the presence of valley-Zeeman term. We will explicitly consider the states with filling vector $(1,1,\nu_3,\nu_4)$. Now we need to specify both the fully filled spinors $\{ \ket{\chi_1},\ket{\chi_2} \}$ and fractionally filled spinors $\{ \ket{\chi_3},\ket{\chi_4} \}$. Again after taking the occupied spinors to be in valley-spin tensor product states and enforcing their orthogonality, the states that minimise the anisotropy energy in Eq.\eqref{Eq:Anisotropy} fall into two classes: spin-ordered and valley-ordered states. The spin-ordered states have the form:
\begin{eqnarray}
    \ket{\chi_1}=\ket{\vec{n}, \vec{s}_3},  \quad \ket{\chi_2}=\ket{-\vec{n}, \vec{s}_4} \nonumber \\
    \ket{\chi_3}=\ket{\vec{n}, -\vec{s}_3},  \quad \ket{\chi_4}=\ket{-\vec{n},-\vec{s}_4} \nonumber
\end{eqnarray}
And their anisotropy energy can be obtained from Eq.\eqref{Eq:Anisotropy} as:
\begin{align}
E_{so}&=-u_{\perp}(1-\nu)\vec{s}_3 \cdot \vec{s_4}-u_z-u_{\perp}(1+\nu) \nonumber \\ &-h((1-\nu_3)s_{3z}+(1-\nu_4)s_{4z})-\Delta(\nu_3-\nu_4).
\end{align}
The valley-ordered states have the spinors:
\begin{eqnarray}
    \ket{\chi_1}=\ket{\vec{n}_3, \vec{s}},  \quad \ket{\chi_2}=\ket{\vec{n}_4, -\vec{s}} \nonumber \\
    \ket{\chi_3}=\ket{-\vec{n}_3, \vec{s}},  \quad \ket{\chi_4}=\ket{-\vec{n}_4,-\vec{s}} \nonumber
\end{eqnarray}
 
 And their anisotropy energy is obtained from Eq. \eqref{Eq:Anisotropy} as:
 \begin{align}
    E_{vo} &=& u_{\perp}(1-\nu)u_{\perp} {\vec n}_ {3\perp} \cdot {\vec n}_ {4\perp} + (1-\nu) u_z n_{3z}n_{4z} -2 \nu u_{\perp}   \nonumber\\
    &-& \nu u_{z}-\Delta( (1-\nu_3)n_{3z}+(1-\nu_4)n_{4z})-h(\nu_3-\nu_4),
\end{align}

\noindent where $\nu=\nu_3+\nu_4$ and ${\vec n}_ {3\perp} \cdot {\vec n}_ {4\perp} =n_{3x} n_{4x}+n_{3y} n_{4y}$. Notice that in the anisotropy energy of valley ordered states with fractionally filled components the occupied spinors have different couplings to the sublattice symmetry breaking strength ($\Delta$), except in the special case when $\nu_3=\nu_4$. This special case includes the integer quantum Hall ferromagnetic states at neutrality ($\nu_3=\nu_4=0$). Therefore generically we expect that the fractional quantum Hall Kekule states will have spinors with different orientations in the valley Bloch sphere, and we dub these states ``Canted Kekule States". For the specific case of fractional fillings $\nu_3\neq \nu_4$, within the anti-ferromagnetic order, there is a competition between a canted AFM (CaAFM)(where the spinors have different orientation in spin Bloch sphere) and a collinear AFM (CoAFM).

\begin{figure*}[]
\includegraphics[width=0.4\textwidth]{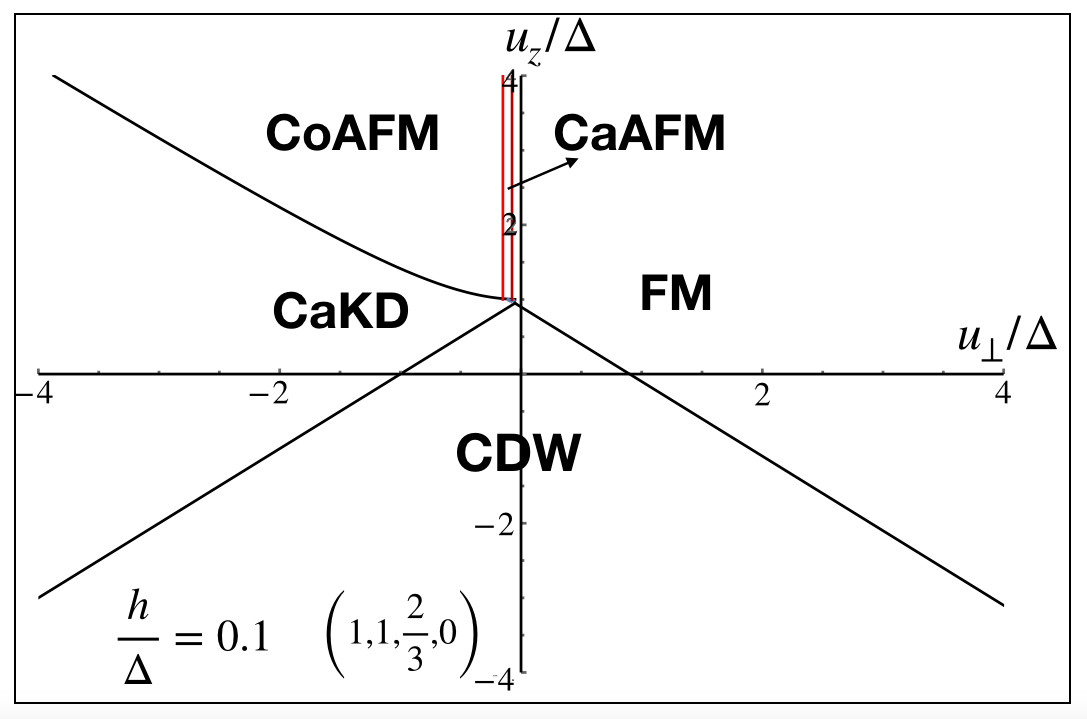}
\includegraphics[width=0.4\textwidth]{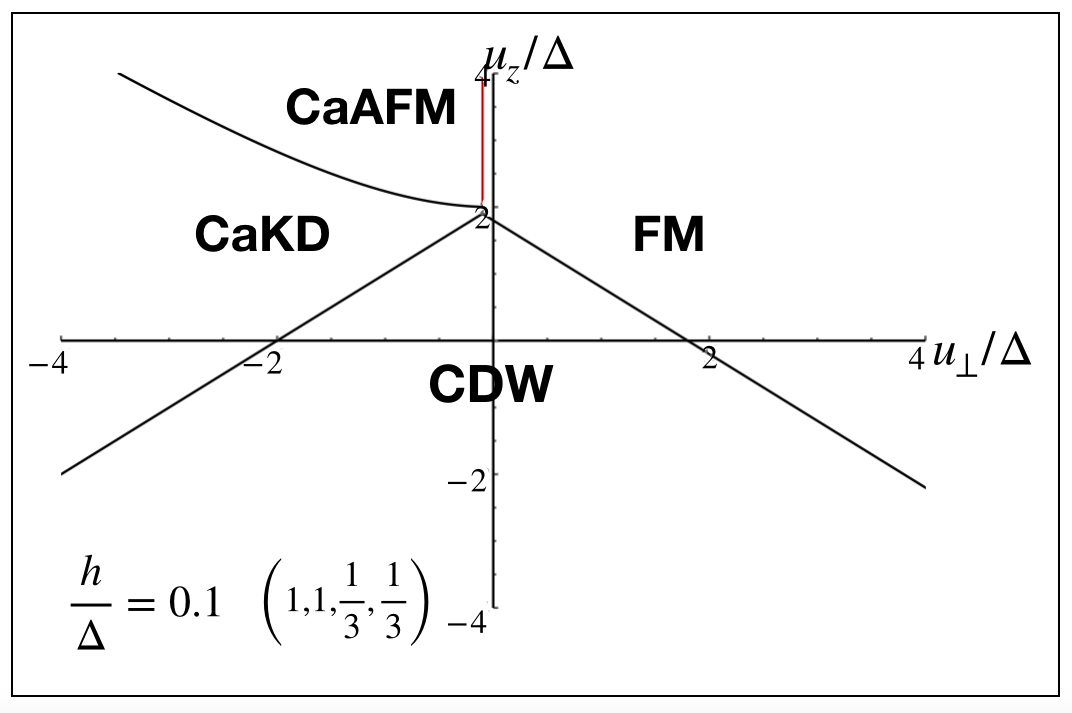}
\caption{Phase diagrams in the presence of both valley-Zeeman and spin-Zeeman terms. The Zeeman and valley-Zeeman field ratio $h/\Delta=0.1$. In the $(1,1,2/3,0)$ state there is a small sliver of region of stability of the CaAFM state. In the $(1,1,1/3,1/3)$ state, the boundary of the CaAFM state with the CoAFM state is pushed to infinity.  }
 \label{fig:PhaseDiagrams}
\end{figure*}

\begin{figure}
    %\centering
  \includegraphics[width=0.4\textwidth]{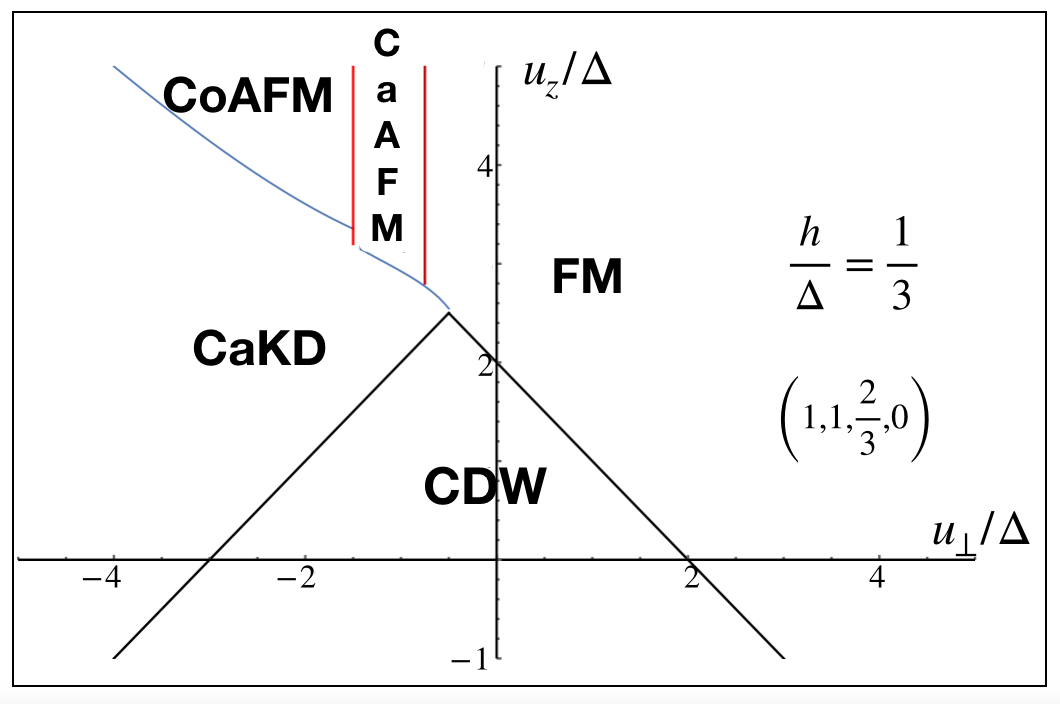}
    \caption{Phase diagram of $(1,1,2/3,0)$ state with Zeeman and valley-Zeeman field ratio $h/\Delta=1/3$.  As the Zeeman field increases the region of stability of CaAFM phase enlargens, but still occupies a narrow strip in the phase diagram.}
    \label{fig:23hbig}
\end{figure}

Minimisation of the energy functionals lead to five distinct phases as shown in Table. \ref{frac-tsymm} and phase diagram for different cases are given in Figs. \ref{fig:PhaseDiagram},\ref{fig:PhaseDiagrams},\ref{fig:23hbig}. 
%The phase boundaries between various phases for $\nu_3=\nu_4$ are obtained by comparing the energies. 
% \begin{itemize}
%     \item CaAFM and FM: $u_{\perp}=-\frac{h(1-\tilde{\nu})}{2(1-2\tilde{\nu})}$
%     \item FM and CDW: $u_z=-u_{\perp}-\frac{(h-\Delta)(1-\tilde{\nu})}{(1-2\tilde{\nu})}$
%     \item CDW and CaKD: $u_z=u_{\perp}+\frac{\Delta(1-\tilde{\nu})}{(1-2\tilde{\nu})}$
%     \item CaKD and CaAFM: $u_z^2-u_{\perp}^2-\frac{(1-\tilde{\nu})^2}{(1-2\tilde{\nu})^2}\Delta^2-\frac{(1-\tilde{\nu})^2}{(1-2\tilde{\nu})^2}\frac{h^2}{2u_{\perp}}(u_z-u_{\perp})=0$
% \end{itemize}
In appendix \ref{App:TwoComp} we describe a mapping from states $(1,\nu,0,0)$ to $(1,1,\nu_3,\nu_4)$ that allows to compute the energies and phase diagrams for other states from the results presented here. Fig. \ref{fig:PhaseDiagram} shows the difference between the phase boundaries in the two cases $\nu_3=2/3 ,\nu_4=0$ and $\nu_3=\nu_4=1/3$, in the absence of Zeeman term $h$. The phase diagrams in the presence of Zeeman term $h$ are shown in Figs.\ref{fig:PhaseDiagrams},\ref{fig:23hbig}. One can see the stark difference between phase boundaries and phases realised for $(1,1,2/3,0)$ and $(1,1,1/3,1/3)$. In the following section, we highlight the differences in these phase diagrams and their relevance for experiments.

\section{Isospin and valley-/spin-order transitions and relation to experiments.}
\label{sec:Transitions}
 We will now describe the range of parameters and the phase diagrams that are accessible in experiments by tuning the magnetic field for the case of graphene aligned with an hBN substrate. Recent experiments have reported the detection of such transitions using magnon transmission \cite{Zibrov18, Zhou2021}. In order to relate to these experiments, we need to express the relation between the experimentally tunable parameters and the parameters of the model. 

The Zeeman term has a linear dependence on the total applied magnetic field $h=2 \mu_B B$, $\mu_B$ is the Bohr magneton, $B =\sqrt{B_{\parallel}^2+B_{\perp}^2}$  (we take the g-factor $g = 2$). We parametrize the coefficients of the anisotropic delta-function interactions, $u_{\perp,z}$, as follows $(u_{\perp},u_z)=g(\cos(\theta),\sin(\theta))$. Typical values for these parameters, obtained from experiments in Refs.\onlinecite{Zibrov18,Zhou2021} are listed in Table \ref{Table:Parameters}.

% The total energy estimate at a certain filling and a given phase is given by :
% \begin{equation}
%     E(\nu)=-\Delta T_z - h S_z - \frac{e^2}{\epsilon \ell_{1T}}\sqrt{B_{\perp}(T)} C(\nu)+ E_a,
%     \label{Eq:Energy}
% \end{equation}
%  \begin{equation}
%     E_a = A_{\perp} u_{\perp} +A_z u_z,
% \end{equation}
The total energy of a state will contain in addition to the anisotropy energies listed in Tables \ref{frac-tsymm}, an SU(4) invariant long-range part of the Coulomb interaction, and thus can be written as:
%  \begin{align}
%   E&= -\Delta  T_z -2\mu_BB_T S_z - \frac{e^2}{\epsilon \ell_{1T}}   C(\nu) \sqrt{B_{\perp}(T)} \\ &+ A_{\perp}(g \cos{\theta})\frac{a}{\ell_{1T}} \frac{e^2}{\epsilon \ell_{1T}} B_T + A_{z}(g \sin{\theta})\frac{a}{\ell_{1T}} \frac{e^2}{\epsilon \ell_{1T}} B_T,
%  \end{align}
\begin{equation}\label{Etot}
    \frac{E}{N_\phi}= E_{\text{ani}} + \frac{e^2}{\epsilon \ell_{1T}}   C(\nu) \sqrt{B_{\perp}(T)},
\end{equation}
  
 Here $\ell_{1T}$ is the magnetic length at 1 Tesla and $E_{\text{ani}}$ is the corresponding  anisotropy energy of the state in question that can be read from Table \ref{frac-tsymm}. In the above we imagine that electrons interact via the ideal  Coulomb interaction $e^2/\epsilon r$ (i.e. not screened by gates). The screening constant, $\epsilon$, accounts both for the screening from the dielectric substrate  as well as the self-screening of graphene, which when accounted at the RPA level leads~\cite{Fogler12} to $\epsilon = \epsilon_{hBN}+\frac{\pi}{2}\alpha$ (with $\alpha=c/(137 v_F) \approx 2.2$). For fully hBN encapsulated samples we take $\epsilon_{hBN}=4.4$. $C$ in Eq.~\eqref{Etot} is a dimensionless constant measuring the ideal Coulomb energy per flux quantum of the FQHE state in consideration, and which we extract from previous exact diagonalization studies~\cite{Xie89,Nielmel00,Vyborny09,Davenport12}.

While the Coulomb energy will not affect the SU(4) spinor orientation of a FQH State, it can play a crucial role determining the competition among different FQH states with different fillings of the flavors. To illustrate this we discuss in detail the competition of the Laughlin-like state and a singlet state at $\nu=2/3$, which has lower Coulomb energy. Their competition has been widely studied specially in two-component systems in the past\cite{Nayak95,Nayak95A, McDonald96,Lay97,Wu12}. The full four-component filling vectors of the states in question are:
\begin{align*}
    \text{Laughlin-like:}&(1,1,\frac{2}{3},0) \\  &\text{vs} \\ \text{Singlet-like:}& (1,1,\frac{1}{3},\frac{1}{3})
\end{align*}
Their Coulomb energy difference per flux quantum obtained from exact diagonalisation studies \cite{Xie89,Nielmel00,Vyborny09,Davenport12}:
\begin{align}\label{E2/3}
    \frac{\delta E^{\rm Coul}_{2/3}}{N_\Phi} \approx 0.006 \frac{e^2}{\epsilon \ell_{1T}} \sqrt{B_{\perp}(T)} 
\end{align}

% These  two states are in fact obtained from a global particle-hole conjugation of the states $(1,1/3)$ and $(2/3,2/3)$ at filling $\tilde{\nu}=4/3$ from the empty $N=0$ Landau level of graphene. These two-component states can further be considered as two component particle-hole conjugation of states $(2/3,0)$ and $(1/3,1/3)$ at $\tilde{\nu}=2/3$. 
 
\begin{table}[ht]
\centering
\begin{tabular}[t]{|c|c|}
\hline
Parameters & Values in experiment at 1T \cite{Zibrov18,Zhou2021}\\ \hline
Valley-Zeeman $\Delta$ & 3.7meV    \\ 
Zeeman $h$	& 0.115meV   \\
$(u_z,u_{\perp}) = g(\cos \theta,\sin \theta)$	&  $g$=3.87meV \\
$\theta$	&  $123 \deg$   \\
$e^2/(\epsilon \ell_{1T})$ & 7.1meV \\
%$g=18 \times 0.215meV $
\hline
\end{tabular}
\caption{List experimental parameters.}
\label{Table:Parameters}
\end{table}

\begin{figure*}[]
\centering
\includegraphics[width=0.6\textwidth]{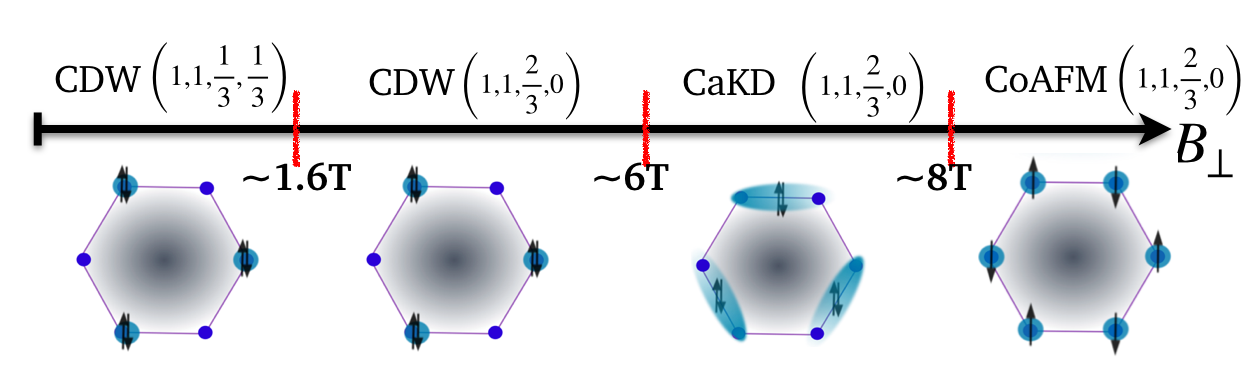}
\includegraphics[width=0.45\textwidth]{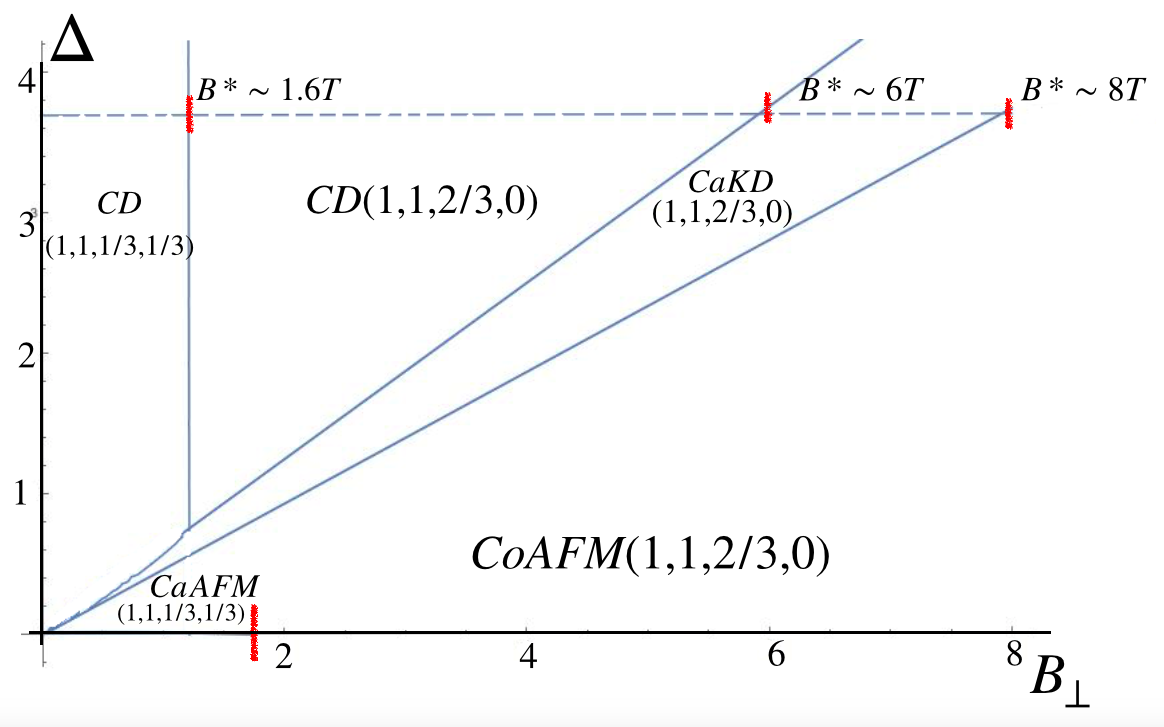}
\includegraphics[width=0.45\textwidth]{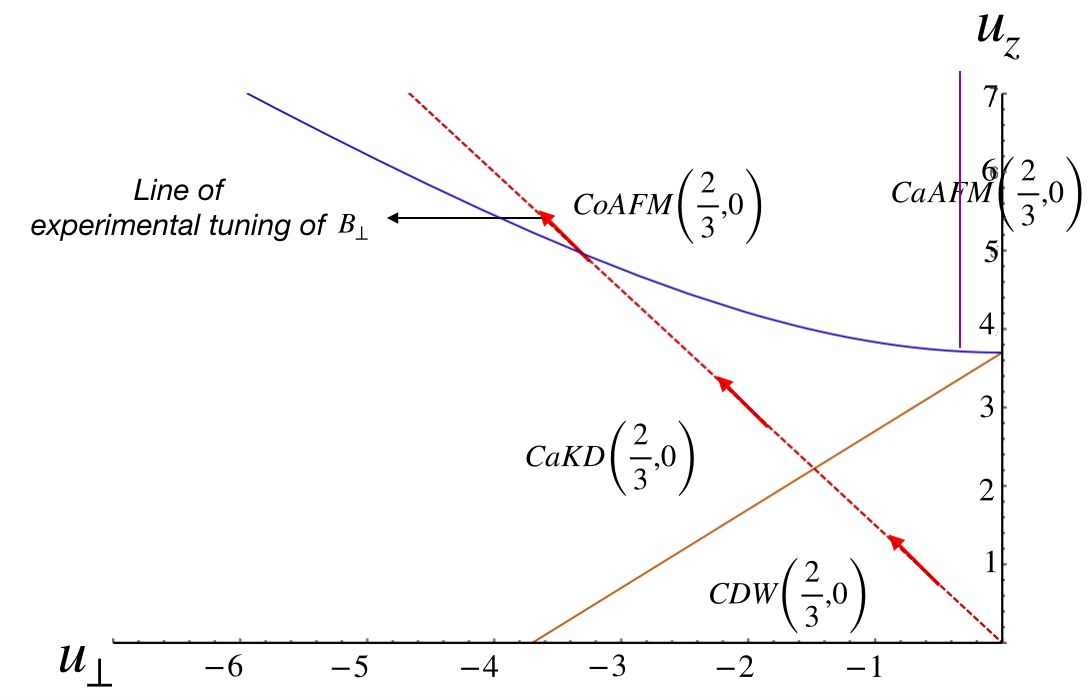}
\caption{(Top) Schematics for different phases and the transitions at total filling $\nu=2/3$ (see also Fig.\ref{fig:AllTransitions}). (Left) Phase diagram as a function of experimental parameters, the valley-Zeeman term $\Delta$ (hBN potential) and the perpendicular magnetic field $B_{\perp}$. The horizontal dotted line indicates a typical value realized in the experiments of Ref.~\onlinecite{Zhou2021}.  (Right) The phase diagram schematic indicating the locus of the system in the $(u_z,u_\perp)$ phase diagram under experimental tuning of magnetic field as shown by dotted line(red). The phase diagram is in the limit of Zeeman field $h$ much smaller than $\Delta$($h_{1T} \sim 0.1 meV, \Delta\sim 3meV$). The region of CaAFM(2/3,0) has been scaled up for visibility.  }
 \label{fig:Transitions}
\end{figure*}

 We estimate from the experiments in Ref.\onlinecite{Zhou2021,Zibrov18} that by tuning the perpendicular field the system follows the line in the $(u_z,u_{\perp})$ space shown in Fig.\ref{fig:Transitions}. Below we describe our key results and predictions.

%\item  Within the framework of our formalism, we provide an estimate of the `critical field' of the CD-CoAFM transition. The behaviour of the total energy as the system goes from CD to AF phase is given in Fig. \ref{fig:AFTransition}. The estimated `critical field' for the transition of the polarised state is $B^*\sim 3.6T$, which is in the ballpark of the experimental value. For higher values of valley-Zeeman(say 6meV), the transition field increases ($B* \sim 6T$).

\subsection{Halperin CDW to Laughlin-like CDW iso-spin transition.}

At low fields, the system is in a CDW state favored by the hBN induced valley Zeeman term $\Delta$ (sublattice staggered potential). However, there is a competition between two distinct CDW states with different fractional occupations. Namely, one CDW state of Laughlin-like character, with filling vector $(1,1,2/3,0)$, and the singlet-like CDW with fillings $(1,1,1/3,1/3)$. These CDW states have the same valley/sublattice polarization but different spin polarization  (their anisotropy energy difference depends only on the Zeeman term). The singlet-like CDW is spin unpolarized and is the ground state at low fields where the Coulomb energy is dominant, while the Laughlin-like CDW becomes the ground state for fields above $\sim 1.6T$ (see Figs.~\ref{fig:AllTransitions} and~\ref{fig:Transitions}). This transition is analogous of those in GaAs \cite{Eisenstein90,Lay97,Cho98,Nielmel00,Feldman09,Feldman12,Peterson15,YuheZhang16} and it features and interesting interplay of symmetry enriched fractionalization as we further discuss in Sec.\ref{sec:TopOrder}.

We emphasize that while this transition should be present in the system, it has so far not been reported in experiments~\cite{Polshyn18,Zhou2021}. A possible reason for this might be that it is realized at small fields where disorder effects can be more important. We hope that future studies can focus on this field range to shed more light on this transition.

%energy difference between the two states is dependant only on the Zeeman term ($ \sim B$) and the Coulomb term($\sim \sqrt{B}$).

%Here we point to a `hidden transition' from a singlet state to a polarised state, that are inaccessible to the regime of parameters under consideration.

 %One expects the singlet $(1,1,1/3,1/3)$ state to win at very low fields, where the Coulomb term is dominant. For the given experimental parameter regime, we find that a transition occurs at  small fields ($\sim 1.6T$) to the Laughlin-like state $(1,1,2/3,0)$ (Fig.\ref{fig:Transitions} and Fig. \ref{fig:AllTransitions}). The tilted field magneto-transport measurements report that Laughlin-like states are measured in the experimental samples \cite{Polshyn18,Zhou2021}. Even though not accessible in the current experimental set-ups, this transition is the counter part of those found in GaAs samples \cite{Eisenstein90,Lay97,Cho98,Nielmel00,Feldman09,Feldman12,Peterson15,YuheZhang16} and are fundamentally important as we explain in Sec.\ref{sec:TopOrder}.

%As mentioned before, at low fields the Laughlin-like state is in a CDW state favored by the hBN induced valley Zeeman term $\Delta$ (sublattice staggered potential)\cite{Polshyn18,Zhou2021}. 

\subsection{The intermediate CaKD phase and transition to CoAFM.} 
One of the key predictions from  our analysis is the occurrence of an intermediate CaKD phase between the CDW and AFM phases, as seen in Fig.\ref{fig:Transitions}.  Notice that unlike the CaKD state realized for the integer quantum Hall ferromagnet at neutrality (Sec.~\ref{sec:Neutrality}), in the fractionally filled case the spinors have different orientations in the valley Bloch sphere. This arises from the unequal coupling of the valley-Zeeman term to the fully filled and fractionally filled spinors, leading to unequal canting angles of the spinors in the valley Bloch sphere. This is why we have termed this state a ``CaKD" (Canted Kekule Distortion) to emphasize this difference from the  KD state at neutrality where the two occupied spinors have a common orientation in the valley Bloch sphere and thus a valley ferromagnetic-like character. Recent tunneling experiments have demonstrated clear and tantalizing evidence for the presence of these states~\cite{liu2021visualizing,coissard2021}.

We estimate the transition from CDW to CaKD phase to occur around $B \sim 6T$.  This transition should be ``dark" within Magnon transmission experiments since there is no change of the spin order, explaining why it was not reported in Refs.~\cite{Zibrov18,Zhou2021}. However, more recently, remarkably beautiful scanning tunneling spectroscopy (STS) experiments have in fact detected a transition from the CDW states to a KD state at neutrality ~\cite{liu2021visualizing,coissard2021}, and therefore these experiments have a great potential to detect the corresponding transition in the fractionally filled case that we predict here. We note in passing that the interesting STS study of Ref.~\onlinecite{PhysRevB.100.085437} also reported the observation of the Kekule state at neutrality in samples of multilayer graphene without the hBN substrate. Here the top layer was viewed as decoupled from the rest, but there was no independent control of the magnetic field and the density of this layer.

%, though the nature of the valley-ordered state was undetermined and the AFM phase is identified with CaAFM based on considerations of phase diagram at neutrality

As the applied field is further increased the CaKD Laughlin-like $(1,1,2/3,0)$ state eventually transitions into a CoAFM phase. We estimate the field of the transition to be $B\sim 8T$. We interpret this as the transition from a valley-ordered phase to an anti-ferromagnetic that has been detected through magnon transmission experiments \cite{Zhou2021}.

There is one additional phase, the CaAFM (Canted AFM) which in the case of the Laughlin-like filling $(1,1,2/3,0)$, occupies a very small region in the parameter space, as in seen in Fig. \ref{fig:PhaseDiagrams}(cf. Also see Fig.\ref{fig:23hbig}). Its region of stability is given by\cite{Sodemann14}:
\begin{equation}
    \frac{(1-\nu_3)(1-\nu_4)}{(1-\nu)(\nu_3-\nu_4)}\geq \frac{u_{\perp}}{h} \geq \frac{(1-\nu_3)(1-\nu_4)}{(1-\nu)(2-\nu)}.
    \label{Eq:stability}
\end{equation}

Because of the smallness of this region, we believe that the experimentally accessible line of parameters for the filling $\tilde{\nu}=2+2/3$ does not cross this phase (see Fig.~\ref{fig:Transitions}).

\section{On the universal topological distinction between the Laughlin-like and singlet-like states.}
\label{sec:TopOrder}
All the transitions occurring among the states with filling vector (1,1,2/3,0) that were discussed previously (see Figs.\ref{fig:AllTransitions} and \ref{fig:Transitions} for a summary) can be understood as a re-orientation of the valley-spin degrees of freedom of the state. However the transition between the (1,1,1/3,1/3) and (1,1,2/3,0)  occurring around $B_{\perp} \sim 1.6T$ is of a fundamentally different kind where a change of the more intrinsic nature of the correlations occurs. The nature of this transition has been the subject of dedicated study of FQH researchers for decades specially in the context of two-component FQH systems\cite{Nayak95,Nayak95A, McDonald96,Lay97,Wu12}. In this section we would like to briefly review the theoretical underpinnings on why this transition is special. We believe that graphene is an unprecedented physical platform that could open new windows to investigate this amazing physics. For simplicity only in this section we will concentrate on these states viewed as two component states, namely as $(2/3,0)$ and $(1/3,1/3)$.

Both the Laughlin like-state $(2/3,0)$ and the singlet-like state $(1/3,1/3)$ possess anyon excitations with same fractional Abelian statistics, namely under exchange of two of them the many-body wave-function acquires phases quantized in units of $\pi/3$. In fact, as detailed in Appendix \ref{App:ChernSimons}, these two states have the same ``topological order"~\cite{WenBook}. This means that if all the symmetries of the problem are ignored, then it is possible to find an adiabatic path smoothly deforming a Hamiltonian with ground state (2/3,0) into a Hamiltonian with ground state (1/3,1/3) without encountering a phase transition. As we will see however, such path does not exist if certain symmetries are enforced, and therefore these states possess distinct ``symmetry enriched topological order". 

But which symmetries need to be enforced to distinguish these states as distinct phases of matter? These states have anyons with charged quantized in units of $e/3$ and charge conservation is not enough to distinguish these two states from a universal point of view. It turns out, however, that these two states have different topological ``spin vectors"~\cite{WenZee,Nayak95,Nayak95A, McDonald96} (see Appendix \ref{App:ChernSimons} for details). Therefore, interestingly, space isotropy and its associated rotational symmetry, should be enough to distinguish these as topologically distinct states, even if the separate conservation of particle number of the two components is broken (e.g. layer number conservation or spin conservation depending on the context).

%The sphere has an $O(3)$ symmetry and particularly the polarised and the singlet state would be distinguished by the $SU(2)$ rotations of the sphere (which is distinct than the internal $SU(2)$ spin or valley-symmetry).

Moreover these Laughlin-like and singlet-like states can be also be distinguished by the properties of their edges. The unpolarised state displays spin-charge separation at the edge in having counter-propagating spin and charge edge modes\cite{Wu12}. Interestingly the presence of separate particle/spin conservation symmetries in the two components leads to the appearance of neutral upstream modes at the interface of Laughlin-like and singlet-like phases\cite{MacDonald90,Kane94,Kane95,Moore97,Hu08,Wang13}. There are proposals for using these modes as building elements for topological quantum computation applications~\cite{Mong14}. These neutral modes have been extensively probed in experiments as well \cite{Stern04,Bid10,Gross12,Gurman12,Ronen18,Fabien19,Wang2021}. We hope that our study can help guide future efforts to further investigate the fascinating physics of these states in graphene.

%Experiments in single layer high density 2DEGs have reported transition at $\nu=2/3,2/5$ between two-component Laughlin-like and singlet-like state \cite{Eisenstein90,Cho98,Kang97,Feldman13} effected through methods involving tilted magnetic fields, pressure induce reduction of g-factor. Possible application of these experimental methods to detect the proposed transition in graphene systems needs to investigated. 
%This was observed through dramatic hysteresis in up and down sweeps of magnetic field.

\section{Summary and Discussion}
\label{sec:Discussion}

We have elaborated on a theory that generalizes the theory of integer quantum Hall ferromagnets to the fractionally filled states and allows to quantitatively predict the patterns of symmetry breaking in multi-component fractional quantum Hall systems, such as graphene. We have in particular exploited this framework to investigate the role of the staggered sub-lattice potential that results from the alignment with a boron nitride substrate on the fractional quantum Hall states in graphene. As a case study, we have made specific predictions for the states realized at a total filling of $\nu=2/3$ above charge neutrality. Within the typical parameter regimes accessed in the experiments, we predict a rich sequence of phase transitions as a function of perpendicular magnetic field, starting from a singlet-like CDW state, then to a Laughlin-like CDW state, then to a Canted Kekule  (CaKD) Laughlin-like state and finally to a collinear Anti-ferromagnet (CoAFM) Laughlin-like state. 

One of our key predictions is the presence of a Canted Kekule (CaKD) Laughlin-like state, in which the occupied spinors point in different directions in the valley Bloch sphere, in contrast to the Kekule state (KD) realized at the integer filling corresponding to neutrality, where the spinors point into a common direction in the Bloch sphere (this state has been detected in very recent experiments \cite{liu2021visualizing,coissard2021}). Another important prediction is the existence of a transition around $B\sim 1.6T$ from a  singlet-like CDW phase to a Laughlin-like CDW state. This transition is hitherto unobserved in graphene, which could offer a new platform to investigate the interesting physics\cite{Nayak95,Nayak95A,McDonald96,Crepel18,Liang19} of the competing singlet-like and Laughlin like states at $2/3$.

We would like to also comment on some of the challenges in relating theory and experiment, which arise partly because some of the orders described here are hard to detect. Recently an ingenious set of experimental methods has detected the spin-ordering transitions through non-local magnon excitation/transmission devices~\cite{Stepanov_2018,Wei18,zhou2020solids,Zhou2021}, but these experiments make hard to detect when the transition involves primarily some change of the valley ordering. However, it has been proposed~\cite{Wei_2021} that in graphene aligned with hBN, the interface of the $\nu=1$ and $\nu=-1$ could be used as a tunable valley-wave source. This could open up avenues to probe various low-energy excitations in the broken valley-/spin-order phases on FQHEs of graphene.

%~\cite{Zhou2021}, has utilized non-local long-range magnon transmission and detection of spin waves/magnons. This was achieved in quantum Hall ferromagnetic phases of graphene \cite{Wei18, Stepanov_2018} through out-of equilibrium occupation of QH edge channels with opposite spin polarisation thus creating a `chemical potential' for generation and detection of magnons. It would be very useful to develop analogous `valley-wave' generation methods that could allow to probe the changes of valley-ordering. 

%Such valley sources, sinks and valley-splitters have already been reported and shown to be operatable as tunable Mach-Zehnder interferometer.

Another interesting avenue for experimentally probing these phases, that has been developed very recently, is the  use of scanning tunneling spectroscopy(STS) to detect and distinguish valley-symmetry broken phases and their transitions~\cite{liu2021visualizing,coissard2021}. It is possible that in the future these probes could allow to pinpoint the presence of the Canted Kekule state that we have found. They are also very powerful for imaging in detail the defects, quasi-particles, domain walls and other interesting textures that could appear in the different competing phases~\cite{feldman2016observation,ding2017imaging,papic2018imaging,tam2020local}.

%at . More recent work \onlinecite{Le2021} has explored more exotic states such as the three-component  FQH states(also studied earlier in \onlinecite{Wu15}). 

%.  appeared and reported the detection of CDW-KD transition and mapping of valley-texture of skyrmionic excitations in the KD phase at charge neutrality,

\begin{acknowledgements}
We thank Haoxin Zhou and Andrea Young for stimulating discussions, and Allan H. MacDonald for previous collaborations that were key stepping stones for this work.
\end{acknowledgements}

\bibliography{main}

\appendix

\section{ Phase diagram with filling $(1,\nu,0,0)$}
 \label{App:TwoComp}
 We want to obtain the phase diagram with fractional multicomponent fillings such as $(1,\nu,0,0)$ and in the presence of valley-Zeeman term. Here we consider only states that are valley ordered. The treatment of spin-active states are not altered in the presence of the valley-Zeeman term and is the same as given in previous works \cite{Sodemann14, Kharitonov12}.

The weighed projection operators (density matrices) for the fully filled and partially filled spinors are given by:
\begin{eqnarray}
P_i= \ket{\chi_1}\bra{\chi_1} \\
P_f= \nu \ket{\chi_f}\bra{\chi_f}.
\end{eqnarray}
For the fillings we consider here $(1,\nu)$, the first component $\chi_1$ is fully filled and the second spinor is partially filled at the fraction $\nu$.

We make the following choices for the spinors
\begin{eqnarray}
\ket{\chi_1}=\ket{n_1}\otimes \ket{\uparrow} \\
P_1=P_{n_1} \otimes P_s
\end{eqnarray}

\begin{eqnarray}
&\ket{\chi_f}=\ket{n_2}\otimes\ket{\downarrow}\\
&P_f=\nu P_{n_2} \otimes P_{-s}=\nu P_2
\end{eqnarray}

Using these we obtain
\begin{equation}
E_a = \nu u_{\perp}(n_{1x} n_{2x}+n_{1y}n_{2y})+\nu u_z n_{1z}n_{2z}- \Delta (n_{1z}+\nu n_{2z})
\label{Eq:Valley-Energy}
\end{equation}

Now let us consider the value of $E_a$ in different phases. In the CDW phase, 
\begin{equation}
   n_{1z}=n_{2z}; \quad  \epsilon_{CDW}=\nu u_z -\Delta (1+\nu)
\end{equation}

Away from the CDW phase, valley symmetry breaking term tends to cant the valley-isospins by different angles, which would destabilise the KD phase.

{\it Determination of the minimal energy valley-ordered state through the method of Lagrange multipliers:}
 
 We need to obtain the minimum of the energy given in Eq. \ref{Eq:Valley-Energy} for the valley ordered state under the constraint for the valley spinor $n_{\perp}^2+n_z^2=1$, let us consider the Lagrangian function:
  \begin{align}
      \mathcal{L}=\nu u_{\perp} n_{1x}n_{2x}+\nu u_z n_{1z}n_{2z}-\Delta (n_{1z}+\nu n_2z)\\ - \frac{\lambda_1}{2}(n_{1x}^2+n_{1z}^2-1) -\frac{\lambda_2}{2}(n_{2x}^2+n_{2z}^2-1),
  \end{align}
  where $\lambda_1,\lambda_2$ are the Lagrangian multipliers.
 
 The optimum is obtained from:
 \begin{eqnarray}
 \frac{\partial \mathcal{L}}{\partial n_{1x}} &=& \nu u_{\perp} n_{2x}-\lambda_1 n_{1x}=0 \\
 \frac{\partial \mathcal{L}}{\partial n_{2x}} &=& \nu u_{\perp} n_{1x}-\lambda_2 n_{2x}=0 \\
 \frac{\partial \mathcal{L}}{\partial n_{1z}}&=& \nu u_z n_{2z}-\lambda_1 n_{1z}-\Delta=0  \\
 \frac{\partial \mathcal{L}}{\partial n_{2z}}&=& \nu u_z n_{1z}-\lambda_2 n_{2z}-\Delta \nu =0 
 \end{eqnarray}
  
  The equations for x-components give a constrain for the Lagrangian multipliers as
  \begin{eqnarray}
  \frac{n_{1x}}{n_{2x}}=\frac{\nu u_{\perp}}{\lambda_1} \quad \frac{n_{1x}}{n_{2x}}=\frac{\lambda_2}{\nu u_{\perp}}  \\
  \implies \lambda_1\lambda_2 =\nu^2 u_{\perp}^2 .
  \end{eqnarray}
  
  Also multiplying the first and the third equations by 
  $n_{1x}$ and $n_{1z}$ respectively and adding them, we get
  \begin{equation}
      \nu u_{\perp} n_{1x}n_{2x}+\nu u_z n_{1z}n_{2z}-\Delta n_{1z} - \lambda_1(n_{1x}^2+n_{1z}^2)=0.
  \end{equation}
  Using the constraint $n_{1x}^2+n_{1z}^2=1$, we get 
  \begin{equation}
      \lambda_1 =\nu u_{\perp} n_{1x}n_{2x}+\nu u_z n_{1z}n_{2z}-\Delta n_{1z}.
      \label{Eq.Lambda1}
  \end{equation}
  
   Similarly using the second and the fourth equations for the optimum conditions, we obtain the expression for $\lambda_2$:
 
  \begin{equation}
       \lambda_2 =\nu u_{\perp} n_{1x}n_{2x}+\nu u_z n_{1z}n_{2z}-\Delta \nu  n_{2z}.
       \label{Eq.Lambda2}
   \end{equation}

   The z-components of the spinors can be obtained by solving the following equations
   
   \begin{eqnarray}
   \lambda_1 n_{1z}-\nu u_z n_{2z}= -\Delta \\
   \nu u_z n_{1z} -\lambda_2 n_{2z}=\nu \Delta.
   \end{eqnarray}

  \begin{equation}
      n_{1z}=\frac{\lambda_2 \Delta+ \nu^2 u_z \Delta}{\nu^2 (u_z^2-u_{\perp}^2)}, \quad  n_{2z}= \frac{\lambda_1 \nu \Delta+\nu u_z \Delta}{\nu^2 (u_z^2-u_{\perp}^2)}.
  \end{equation}

The Lagrangian multipliers can be determined from the above equations for $n_{ix},n_{iz}$, $\lambda_1 \lambda_2= \nu^2 u_{\perp}^2$ and the constraints $n_{ix}^2+n_{iz}^2=1$.

\begin{align*}
 & n_{1x}^2  +n_{1z}^2=1 \\
 n_{1x}=&\frac{\nu u_{\perp}}{\lambda_1}n_{2x}, \quad n_{2x}^2=1-n_{2z}^2 \\
 \bigg(\frac{\nu u_{\perp}}{\lambda_1}n_{2x}  &\bigg)^2  +\bigg(\frac{\lambda_2 \Delta+ \nu^2 u_z \Delta}{\nu^2 (u_z^2-u_{\perp}^2)}\bigg)^2=1 \\
 \frac{\nu^2u_{\perp}^2}{\lambda_1^2} \bigg(1-& \bigg(\frac{\lambda_1 \nu \Delta+\nu u_z \Delta}{\nu^2 (u_z^2-u_{\perp}^2)} \bigg)^2 \bigg) +\bigg(\frac{\lambda_2 \Delta+ \nu^2 u_z \Delta}{\nu^2 (u_z^2-u_{\perp}^2)}\bigg)^2 =1
\end{align*}

 After some algebra, we obtain,
 \begin{equation}
     \lambda_2 =\pm \nu^2 u_{\perp} \sqrt{\frac{u_z^2-u_{\perp}^2-\Delta^2}{\nu^2 (u_z^2-u_{\perp}^2)-\Delta^2} }
 \end{equation}
 
Given this one can determine the energy minimum. The energy function is given by
\begin{equation}
    E_a = \nu u_{p}(n_{1x} n_{2x}+n_{1y}n_{2y})+\nu u_z n_{1z}n_{2z}- \Delta (n_{1z}+\nu n_{2z})
\end{equation}
From Eq.\ref{Eq.Lambda1} and Eq.\ref{Eq.Lambda2}, we get

\begin{equation}
E_a =\lambda_1-\nu \Delta n_{2z}=\lambda_2 -\Delta n_{1z}
\end{equation}

using the above expression for $\lambda_2$, we get
\begin{eqnarray}
    E_a^* =-\nu^2 u_{\perp} \sqrt{\frac{u_z^2-u_{\perp}^2-\Delta^2}{\nu^2 (u_z^2-u_{\perp}^2)-\Delta^2} } \bigg(1 & -& \frac{\Delta^2}{\nu^2 (u_z^2 -u_{\perp}^2)} \bigg) \nonumber \\& -& \frac{u_z \Delta^2}{u_z^2 -u_{\perp}^2}
    \label{Eq:Analytical}
\end{eqnarray}
one can check that for $\nu=1$, this equation reduces to the minimum energy expression at neutrality.

{\it Phase boundary between CDW and FM phases:}
In the CDW phase, the valley spinor is ordered such that $n_{1z}=n_{2z}=1$. Therefore the equations obtained from the optimisation reduce to
\begin{align}
    \lambda_1 n_{1z}-\nu u_z n_{2z}= -\Delta \implies \lambda_1=\nu u_z -\Delta \\
   \nu u_z -\lambda_2 n_{2z}=\nu \Delta \implies \lambda_2=\nu u_z -\nu \Delta \\
   n_{1z}=1 \implies  \frac{\lambda_2 \Delta+ \nu^2 u_z \Delta}{\nu^2 (u_z^2-u_{\perp}^2)}=1 \\
    n_{2z}=1 \implies  \frac{\lambda_1 \nu \Delta+\nu u_z \Delta}{\nu^2 (u_z^2-u_{\perp}^2)}=1
\end{align}

\begin{figure}
    \centering
    \includegraphics[width=0.4\textwidth]{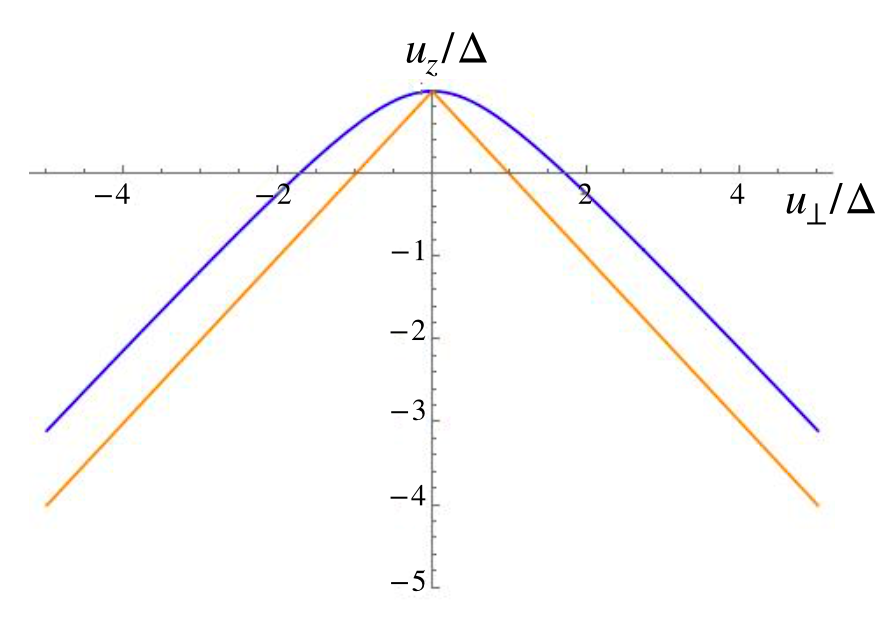}
    \caption{The hyperbola of instability of CDW phase and the phase boundary obtain by comparing the CDW and FM energies}
    \label{fig:CDboundary}
\end{figure}

Solving these equations, one obtains the hyperbola(Fig. \ref{fig:CDboundary}) 
\begin{equation}
    \bigg( u_z- \frac{\Delta(1+\nu)}{2\nu} \bigg)^2 -u_{\perp}^2 =\bigg(\frac{\Delta(1-\nu)}{2\nu} \bigg)^2
\end{equation}

The intercept of the lower branch of the hyperbola with the $u_z$ axis is given by
\begin{eqnarray}
   % u_z &-&\frac{\Delta(1+\nu)}{2\nu} = -\frac{\Delta(1-\nu)}{2\nu} \\
   u_{z0} =\Delta
    \end{eqnarray}
    
Also, comparing the energy of the CDW state with the FM state
\begin{eqnarray}
    E_{CDW}=\nu u_z-\Delta(1+\nu) \\
    E_{FM}=-\nu (2u_{\perp}+u_z)-\Delta(1-\nu) 
\end{eqnarray}.
Comparing these two the phase boundary is given by (Fig. \ref{fig:CDboundary})
\begin{equation}
    u_z=-u_{\perp} +\Delta 
\end{equation}

{\it Phase boundary between caKD and AFM:} 
The phase phase boundary can be obtained by comparing the above expression for $E_a^*$ in Eq.\ref{Eq:Analytical} with the energy for AFM $\epsilon_{AFM}=\nu u_{\perp} -\Delta(1-\nu)$. Analytically, this does not give any simple expression for the phase boundary. Nevertheless, one can see that the expression for $\lambda_2$ is real only for $\nu^2(u_z^2-u_{\perp}^2)>\Delta^2$ and also $u_z^2-u_{\perp}^2> \Delta^2$, beyond, which it becomes complex. Therefore, the minimal energy obtained is valid only below the (upper branch of )hyperbola 
\begin{equation}
    u_z^2-u_{\perp}^2= \Delta^2
\end{equation}
and also above the hyperbola $u_z^2-u_{\perp}^2=\frac{\Delta^2}{\nu^2}$. $\lambda_2$ will again be real for  $\nu^2(u_{\perp}^2-u_z^2)>\Delta^2$ and  $u_{\perp}^2-u_z^2> \Delta^2$.
Here we have considered only states that are valley ordered. The energies of spin-ordered states are not affected by the valley-Zeeman term.

\subsection*{Mapping from $(1,1,\nu_3,\nu_4)$ to $(1,\nu,0,0)$}
 
 The energy of the valley-ordered states for $(1,1,\nu_3,\nu_4)$ can be mapped to the energy of the valley-ordered states of $(1,\nu,0,0)$.  The energy for the latter is given by
 \begin{equation}
E_a = \nu u_{\perp}n_{1x} n_{2x} +\nu u_z n_{1z}n_{2z}- \Delta (n_{1z}+\nu n_{2z})
\end{equation}

Upto constants, the energy for the $(1,1,\nu_3,\nu_4)$ state is
\begin{align*}
  E_a=  u_{\perp}(1-\nu) (n_{3x} n_{4x} +u_z (1-\nu)n_{3z}n_{4z} -2\nu u_{\perp}-\nu u_{z} 
    \\-\Delta( (1-\nu_3)n_{3z}+(1-\nu_4)n_{4z})
\end{align*}
 Now lets define the following scaled quantities:
 \begin{flalign}
     \bar{\nu}=\frac{1-\nu_4}{1-\nu_3}; \quad \bar{u}_{\perp}&=\frac{1-\nu}{1-\nu_4}u_{\perp};\quad \bar{u}_z=\frac{1-\nu}{1-\nu_4}u_z ;\\ \bar{\epsilon}_a &= \frac{E_a}{1-\nu_3}
     \label{rescale}
 \end{flalign}
Upon this rescaling, the energy for $(1,1,\nu_3,\nu_4)$ is now:

  \begin{equation}
\bar{\epsilon}_a = \bar{\nu} \bar{u}_{\perp}n_{1x} n_{2x}+\bar{\nu} \bar{u_z} n_{1z}n_{2z}- \Delta (n_{1z}+\bar{\nu} n_{2z})
\end{equation}
This exactly matches the expression for $(1,\nu,0,0)$ but with scaled quantities.

For the case of $\nu_3=\nu_4$, $\bar{\nu}=1$. Therefore the phase diagram of $(1,1,\nu_3,\nu_3)$ is identical to the one for fully filled state and for $\nu_4=0$, the phase diagram resembles the one for $(1,\nu,0,0)$ state.

\section{Phase diagram with filling $(1,1,\nu_3,\nu_4)$}

\label{App:Lagrange}

Here we will consider the valley-ordered states of type $(1,1,\nu_3,\nu_4)$ in the presence of the valley-Zeeman term to obtain the anisotropy energies and phase boundaries. The anisotropy energy for the $(1,1,\nu_3,\nu_4)$ valley ordered state is 
\begin{eqnarray*}
    E_a = (1-\nu)u_{\perp} n_{3x} n_{4x} + (1-\nu) u_z n_{3z}n_{4z} -2\nu u_{\perp}-\nu u_{z} \\
    -\Delta( (1-\nu_3)n_{3z}+(1-\nu_4)n_{4z}),
\end{eqnarray*}
where $\nu=\nu_3+\nu_4$ and we have put $n_{3y}=n_{4y}=0$ for simplicity.
The CDW phase is characterised by $n_{3z}=n_{4z}=1$. The energy corresponding to that phase is 
\begin{equation}
    E_a=(1-2\nu)u_z-2\nu u_{\perp}-\Delta(2-\nu)
\end{equation}.

The formation of KD phase is obstructed by the valley-Zeeman term. One needs to find out the valley-spinor configuration that minimises the anisotropy energy.  We shall follow the method of Lagrangian multipliers

{\it Lagrangian multipliers for finding the valley-optimal states}
The Lagrangian function  is given by
\begin{align*}
    \mathcal{L}&= u_{\perp}(1-\nu) n_{3x} n_{4x} +u_z (1-\nu)n_{3z}n_{4z}\\ &- 2\nu u_{\perp}-\nu u_{z} 
    -\Delta( (1-\nu_3)n_{3z}+(1-\nu_4)n_{4z})\\ &- \frac{\lambda_1}{2}(n_{3x}^2+n_{3z}^2-1) -\frac{\lambda_2}{2}(n_{4x}^2+n_{4z}^2-1).
\end{align*}

The optimum is obtained from:
 \begin{align}
 \frac{\partial \mathcal{L}}{\partial n_{3x}}= (1-\nu) u_{\perp} n_{4x}-\lambda_1 n_{3x}=0 \\
  \frac{\partial \mathcal{L}}{\partial n_{4x}}= (1-\nu) u_{\perp} n_{3x}-\lambda_2 n_{4x}=0. \\
  \frac{\partial \mathcal{L}}{\partial n_{3z}}= (1-\nu) u_z n_{4z}-\lambda_1 n_{3z}-\Delta (1-\nu_3)=0  \\
  \frac{\partial \mathcal{L}}{\partial n_{4z}}= (1-\nu) u_z n_{3z}-\lambda_2 n_{4z}-\Delta (1-\nu_4) =0 
 \end{align}
  
  The equations for x-components give a constrain for the Lagrangian multipliers as
  \begin{eqnarray}
  \frac{n_{3x}}{n_{4x}}=\frac{(1-\nu) u_{\perp}}{\lambda_1} \quad \frac{n_{3x}}{n_{4x}}=\frac{\lambda_2}{(1-\nu) u_{\perp}}  \\
  \implies \lambda_1\lambda_2 =(1-\nu)^2 u_{\perp}^2 .
  \end{eqnarray}
  
  Multiplying the first and the third equations by 
  $n_{3x}$ and $n_{3z}$ respectively and adding them, we get
  \begin{align*}
      (1-\nu) u_{\perp} n_{3x}n_{4x}+ (1-\nu) u_z n_{3z}n_{4z} &-\Delta (1-\nu_3) n_{3z} \\  - &\lambda_1(n_{3x}^2+n_{3z}^2)=0.
  \end{align*}
  Using the constraint $n_{3x}^2+n_{3z}^2=1$, we get 
  \begin{equation}
      \lambda_1 =(1-\nu) u_{\perp} n_{3x}n_{4x}+(1-\nu)u_z n_{3z}n_{4z}-\Delta (1-\nu_3) n_{3z}.
      \label{}
  \end{equation}
  
   Similarly using the second and the fourth equations for the optimum conditions, we obtain the expression for $\lambda_2$:
   \begin{equation}
       \lambda_2 =(1-\nu) u_{\perp} n_{3x}n_{4x}+(1-\nu) u_z n_{3z}n_{4z}-\Delta (1-\nu_4)  n_{4z}.
       \label{}
   \end{equation}
   
    The z-components of the spinors can be obtained by solving the following equations
   \begin{eqnarray}
   -\lambda_1 n_{3z}+(1-\nu) u_z n_{4z}= \Delta (1-\nu_3) \\
   (1-\nu)u_z n_{3z} -\lambda_2 n_{4z}= \Delta (1-\nu_4).
   \end{eqnarray}
   
  \begin{eqnarray}
  \label{Eq:OptimalValeySpin}
      n_{3z}=\frac{\lambda_2 \Delta(1-\nu_3) + (1-\nu)(1-\nu_4) u_z \Delta}{(1-\nu)^2 (u_z^2-u_{\perp}^2)}  \nonumber \\  n_{4z}= \frac{\lambda_1 (1-\nu_4) \Delta+ (1-\nu)(1-\nu_3) u_z \Delta}{(1-\nu)^2 (u_z^2-u_{\perp}^2)}.
  \end{eqnarray}

The Lagrangian multiplier $\lambda_2$ is obtained to be
 \begin{equation}
     \lambda_2 =\pm (1-\nu) u_{\perp} \sqrt{\frac{(1-\nu)^2(u_z^2-u_{\perp}^2)-  (1-\nu_4)^2\Delta^2}{(1-\nu)^2 (u_z^2-u_{\perp}^2)- (1-\nu_3)^2\Delta^2} }
 \end{equation}
 This is real for $(1-\nu)^2 (u_z^2-u_{\perp}^2)> (1-\nu_3)^2 \Delta^2$ and the phase boundary is given by:
 \begin{equation}
     u_z^2- u_{\perp}^2=\frac{(1-\nu_3)^2}{(1-\nu)^2} \Delta^2
 \end{equation}
 There is another boundary beyond which $\lambda_2$ is a complex number:
 \begin{equation}
     u_z^2- u_{\perp}^2=\frac{(1-\nu_4)^2}{(1-\nu)^2} \Delta^2
 \end{equation}
 For $\nu_3=\nu_4$, the boundaries are the same and are separated for the case of $\nu_3>\nu_4$.

\subsubsection{Boundary of stability of CDW phase}
\begin{eqnarray}
      n_{3z}=\frac{\lambda_2 \Delta(1-\nu_3) + (1-\nu)(1-\nu_4) u_z \Delta}{(1-\nu)^2 (u_z^2-u_{\perp}^2)} \\  n_{4z}= \frac{\lambda_1 (1-\nu_4) \Delta+ (1-\nu)(1-\nu_3) u_z \Delta}{(1-\nu)^2 (u_z^2-u_{\perp}^2)}.
  \end{eqnarray}
  
  In the CDW phase, $n_{3z}=n_{4z}=1$.
  \begin{eqnarray}
      n_{3z}=\frac{\lambda_2 \Delta(1-\nu_3) + (1-\nu)(1-\nu_4) u_z \Delta}{(1-\nu)^2 (u_z^2-u_{\perp}^2)} =1.
  \end{eqnarray}
The Lagrangian multiplier $\lambda_2$ is given by
\begin{equation}
    \lambda_2=(1-\nu) u_z -\Delta(1-\nu_4)
\end{equation}
Plugging these in to the equation for $n_{3z}=1$and simplifying, one gets
\begin{equation}
    u_z=\sqrt{u_{\perp}^2+ \frac{\Delta^2 (\nu^2-4\nu_3 \nu_4)}{4(1-\nu)^2}}+\frac{\Delta}{2} \frac{2-\nu}{1-\nu}
\end{equation}

%  \begin{figure}[]
%   \centering
% \begin{subfigure}[]{0.5\textwidth}
% \includegraphics[width=\textwidth]{CDverylow.jpg}
% \caption{}
% \end{subfigure}
%  \hfill
% \begin{subfigure}[]{0.5\textwidth}
% \includegraphics[width=\textwidth]{CDdifference.jpg}
% \caption{}
% \end{subfigure}
%      \hfill
% \caption{(a)Difference between the energies of Laughlin-like and singlet states within the CDW phase. (a)There is a transition from singlet to Laughlin-like state at extremely small fields($\sim 30mT$) and (b)only the Laughlin-like state persists at experimentally relevant fields.}
%  \label{fig:CDTransition}
% \end{figure}

%  \begin{figure}[]
%   \centering
% \begin{subfigure}[]{0.5\textwidth}
% \includegraphics[width=\textwidth]{aCoAFM.jpg}
% \caption{}
% \end{subfigure}
%  \hfill
% \begin{subfigure}[]{0.5\textwidth}
% \includegraphics[width=\textwidth]{bCoAFM.jpg}
% \caption{}
% \end{subfigure}
%      \hfill
% \caption{Difference between the energies of Laughlin-like with different CoAFM order and singlet states in CaAFM order.(a) Transition from $\alpha CoAFM$ order with total isospins $(T_z,S_z)=(2/3,-2/3)$ to CaAFM at $\sim 28T$. (b)Transition from CaAFM back to Laughlin-like but $\beta CoAFM$ with total isospins $(T_z,S_z)=(-2/3,2/3)$ at $\sim 38T$}
%  \label{fig:CoAFMtransition}
% \end{figure}

\subsubsection{Boundary between FM and CDW phases.}
The energy of the FM phase is given by
\begin{equation}
    \epsilon_{FM}=-u_z -2u_{\perp}-\Delta(\nu_3-\nu_4)
\end{equation}
The energy of the CDW phase is
\begin{equation}
    \epsilon_{CDW}=(1-2\nu)u_z -2\nu u_{\perp} -\Delta(2-\nu)
\end{equation}
The boundary between the phase is obtained by equating the above two expressions:
\begin{equation}
    u_z= -u_{\perp}+\Delta \frac{(1-\nu_3)}{(1-\nu)}
\end{equation}
 The transition to FM phase from CDW phase appears before the boundary of stability of CDW phase.

\section{Topological orders of Multicomponent states.}
\label{App:ChernSimons}

On considering the Chern-Simons description of the Laughlin-like and singlet multi-component states at $2/3$ filling, there is a subtle distinction in the intrinsic nature of the states. This distinction is seen particularly when the states are on a manifold with intrinsic curvature such as a sphere. The Chern-Simons theory for the multi-component FQHEs on a curved manifold is given by  \cite{Wen_1995,fradkin_2013}.

 \begin{align}
     \mathcal{L}=-\frac{1}{4\pi} K_{IJ}\epsilon_{\mu\nu \lambda} a_I^\mu \partial^\nu a_J^\lambda -\frac{e}{2\pi}A_\mu t_I \epsilon_{\mu \nu \lambda} \partial^\nu a^\lambda_I \\
     +s_I \omega^i  \epsilon_{i\nu\lambda}\partial_\nu a^\lambda_I
 \end{align}
    Here, $t_I$ is the charge vector and $s_I$ is the spin vector indicating the charge and spin in each flavors/components rspectively.  $K_{IJ}$ is a $2\times 2$ symmetric matrix(considering 2-component states only here.)
% \begin{equation}
%     K= \begin{pmatrix}
%     m_1 & n \\
%     n & m_2
% \end{pmatrix}
% \end{equation}

A given FQH state is completely characterised by the triplet $(K,\vec{t},\vec{s})$ in the presence of a compact curved manifold. Otherwise, it suffices to specify only $(K_{IJ},\vec{t})$. The $(K,\vec{t},\vec{s})$ are unique upto $SL(2,Z)$ transfromations:
\begin{equation}
    K \rightarrow WKW^T;\quad \vec{t} \rightarrow W \vec{t};\quad \vec{s}\rightarrow W\vec{s}
\end{equation}

The K-matrix for the Laughlin-like state is given by\cite{Wen_1995}:
\begin{equation}
    K=\begin{pmatrix}
    1&2 \\
    2&1
    \end{pmatrix} ;\quad t^T=(1,1);\quad s^T=(1/2,-1/2).
\end{equation}
The state corresponding to this description is known to be obtained from Jain construction\cite{Wen_1995}. This is equivalent to a state obtained from a $\nu=-1/3$ state of holes on a $\nu=1$ quantum Hall state of electrons, for which the K-matrix is given by:
\begin{equation}
    K=\begin{pmatrix}
    1&0 \\
    0&-3
    \end{pmatrix} ;\quad t^T=(1,-1);\quad s^T=(1/2,-3/2).
\end{equation}
The above two states are related by the transformation 

\begin{equation}
W=\begin{pmatrix} 1&0 \\ -2 &1 \end{pmatrix}.
\end{equation}

On the other hand, the singlet state is described by:
\begin{equation}
    K=\begin{pmatrix}
    1&2 \\
    2&1
    \end{pmatrix} ;\quad t^T=(1,1);\quad s^T=(1/2,1/2).
\end{equation}
This is not equivalent to the Laughlin-like state by any $SL(2,Z)$ transformation, because of their different topological spins.

%Thus, the transition we have proposed in the previous sections between the Laughlin-like $(1,1,2/3,0)$ and singlet $(1,1,1/3,1/3)$ states is of this nature. It would be interesting to investigate in future the theory of the boundary between these two phases. There have been some investigations on the possibility of appearance of a non-Fermi liquid effective field theory for a possible finite temperature metallic state at the transition\cite{Nayak95,Nayak95A}.

\end{document}